\documentclass[aps,preprint,groupedaddress]{revtex4-2}
\usepackage{xcolor}
\usepackage{eurosym}
\usepackage{graphicx}
\usepackage{amsbsy, amsmath, amssymb}
\usepackage{hyperref}
\usepackage{ulem}
\usepackage{float}

\begin{document}

\title{Unsupervised topological learning\\ for identification of atomic structures}
\author{S\'ebastien Becker$^{1,2}$, Emilie Devijver$^{2}$, R\'emi Molinier$^{3}$, and No\"{e}l Jakse$^{1}$}
\affiliation{$^{1}$Univ. Grenoble Alpes, CNRS, Grenoble INP, SIMaP, F-38000 Grenoble, France\\$^{2}$Univ. Grenoble Alpes, CNRS, Grenoble INP, LIG, F-38000 Grenoble, France\\$^{3}$Univ. Grenoble Alpes, CNRS, IF, F-38000 Grenoble,France\\}
\date{\today }

\begin{abstract}
We propose an unsupervised learning methodology with descriptors based on Topological Data Analysis (TDA) concepts to describe the local structural properties of materials at the atomic scale. Based only on atomic positions and without \textit{a priori} knowledge, our method allows for an autonomous identification of clusters of atomic structures through a Gaussian mixture model. We apply successfully this approach to the analysis of elemental Zr in the crystalline and liquid states as well as homogeneous nucleation events under deep undercooling conditions This opens the way to deeper and autonomous study of complex phenomena in materials at the atomic scale.
\end{abstract}

\keywords{atomic structures, molecular dynamics, homogeneous nucleation, unsupervised learning, topological data analysis}
\maketitle

\newpage

\section{Introduction}

Knowledge of microscopic events driving properties of matter is of fundamental and technological interest. More precisely, understanding the structure at the atomic scale is crucial for the design of new materials, which can play a role in solving contemporary economical and social issues such as the reduction of energy consumption or new drug design \cite{Sosso2016}. However, detailed information on such a small spatial scale is still out of reach experimentally.

To access physical and chemical properties at the atomic scale, simulation tools such as molecular dynamics (MD) represent dedicated in-depth means \cite{Frenkel1996}. Thanks to the increase of computational power, such simulations can nowadays easily reach millions of atoms evolving through several nanoseconds. Although production of large amounts of valuable data has been streamlined, the processing of the so-called “big data” is far from being obvious and has led to the fourth paradigm \cite{Hey2009} of science in many scientific fields in the last two decades. Likewise, in condensed matter physics and materials science, new protocols and tools based on supervised learning and neural networks \cite{Geiger2013, Dietz2017, Boattini2018, DeFever2019} as well as unsupervised learning \cite{Reinhart2017, Reinhart2018, Ceriotti2019, Boattini2019, Paret2020}, or even combinations of these approaches \cite{Spellings2018} are currently built at an accelerated rate.

In this paper, we propose an unsupervised method to analyze structural information, where descriptors from atomic positions are constructed using persistent homology (PH), a classical TDA tool \cite{Motta2018}. PH has been successfully applied to MD simulations to analyze the medium-range structural environments \cite{Tanaka2019} in amorphous solids \cite{Nakamura2015, Hiraoka2016, Hirata2020}, ice \cite{Hong2019}, and complex molecular liquids \cite{Sasaki2018}. However, to the best of our knowledge, PH has never been used as a descriptor to encode local atomic structures, whereas it provides topological features at different resolutions and for different levels of homology. Similar local atomic structures, with respect to those topological features, are clustered using a Gaussian mixture model (GMM). To illustrate the potential of this approach, we analyze here previous simulations of  homogeneous crystal nucleation of elemental Zirconium (Zr) \cite{Becker2020}.

Section \ref{Section:UnsupervisedLearningBasedOnTopologicalModeling} is dedicated to a  presentation of our method illustrated on pure Zr simulations, while Section \ref{Section:ApplicationApproach} presents and discusses the physical results highlighted on these simulations. Finally, Section \ref{Section:Conclusion} draw the conclusions.

\newpage
\section{Unsupervised learning based on topological descriptors}
\label{Section:UnsupervisedLearningBasedOnTopologicalModeling}
We present in this section the complete methodology behind our unsupervised protocol, giving all the steps necessary for its implementation.

\subsection{Data production}
\label{Section:Dataproduction}

\begin{figure}[h!]
	\centering
	\includegraphics[width=0.6\textwidth]{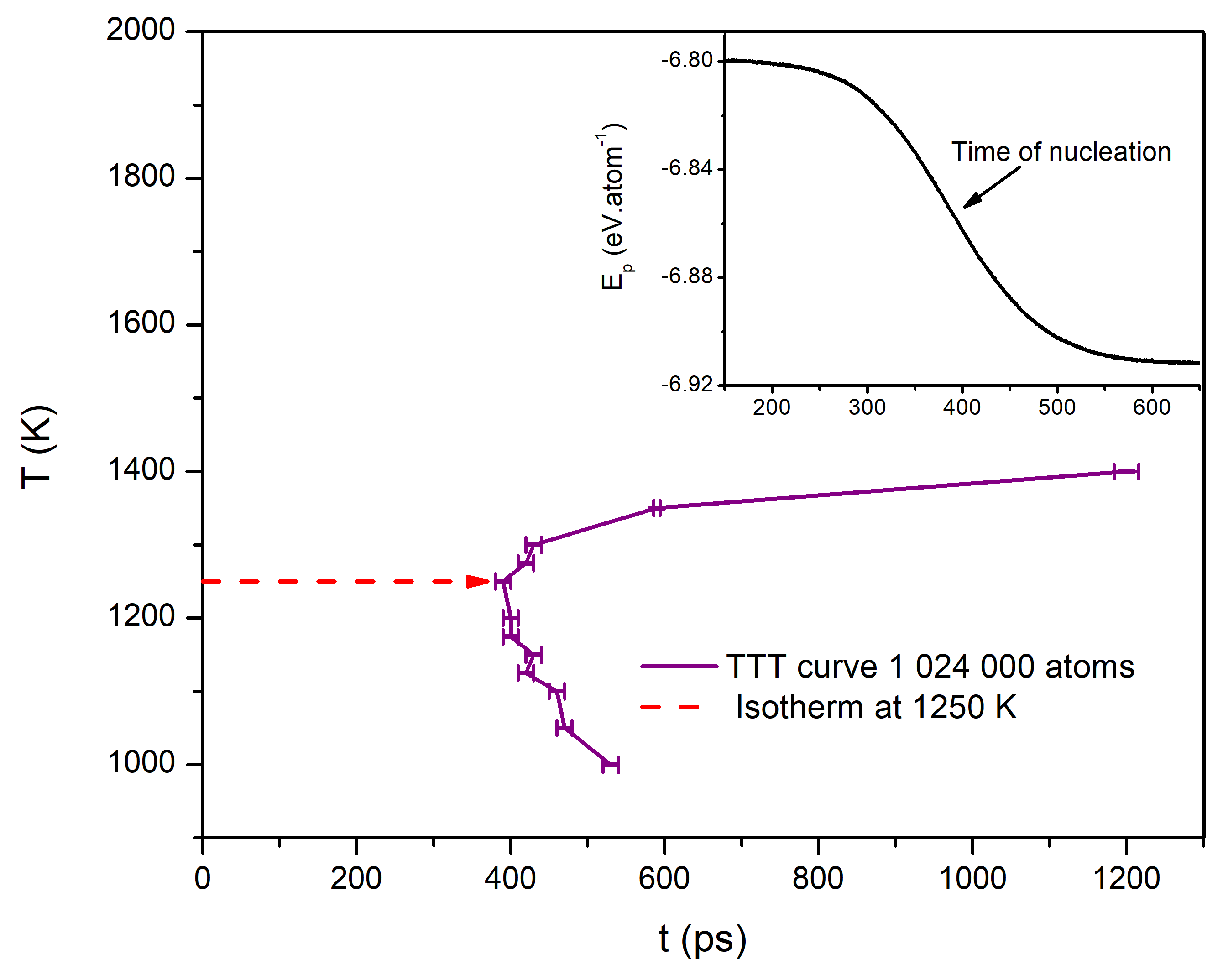}
	\caption{\label{fig:1}Time-temperature-transformation curve with 1 024 000 atoms in a temperature range near the nose, which was obtained over $5$ independent runs for each temperature . Inset: evolution of the potential energy as a function of time along a nucleation process. See more details in \cite{Becker2020}.}
\end{figure}

Our unsupervised learning approach is illustrated on pure Zr simulations. All the simulations mentioned in this paper have been performed with the \textsc{lammps} code \cite{Plimpton1995}, in the isobaric-isothermal ensemble with the Nosé-Hoover thermostat and barostat \cite{Allen2017} and interatomic interaction described by the MEAM potential \cite{Baskes1992} developed in \cite{Becker2020}. For a detailed description of the procedure of the simulation, we refer the reader to Ref.~\cite{Becker2020}, and only the essential characteristics are recalled here. The phase space trajectory is obtained by integrating Newton's equations numerically using the Verlet's algorithm in its velocity form, with a time step of 2 fs. A simulation box with $N=1$ $024$ $000$ atoms is considered with periodic boundary conditions (PBC) in the three directions of space. This system is first equilibrated in the liquid state above the melting points at $T=2500$ K and quenched with a cooling rate of $10^{12}$ K/s at ambient pressure down to the nose of the time-temperature-transformation (TTT) curve at $T=1250$ K as shown in Fig.~\ref{fig:1}. The homogeneous nucleation process is then observed along this isotherm.    

When the onset of nucleation is observed, typically at the nucleation time, a $N$-atom configuration is extracted and its inherent structure is used to build a training set. The inherent structures \cite{Stillinger1982} are obtained by minimizing the energy by means of a conjugate gradient algorithm, implemented in \textsc{lammps}, to bring the system to its local minimum in the potential energy surface to suppress the thermal noise.

\subsection{Data preparation}
\label{Section:DataPreparation}

\begin{figure}[h!]
	\centering
	\includegraphics[width=1\textwidth]{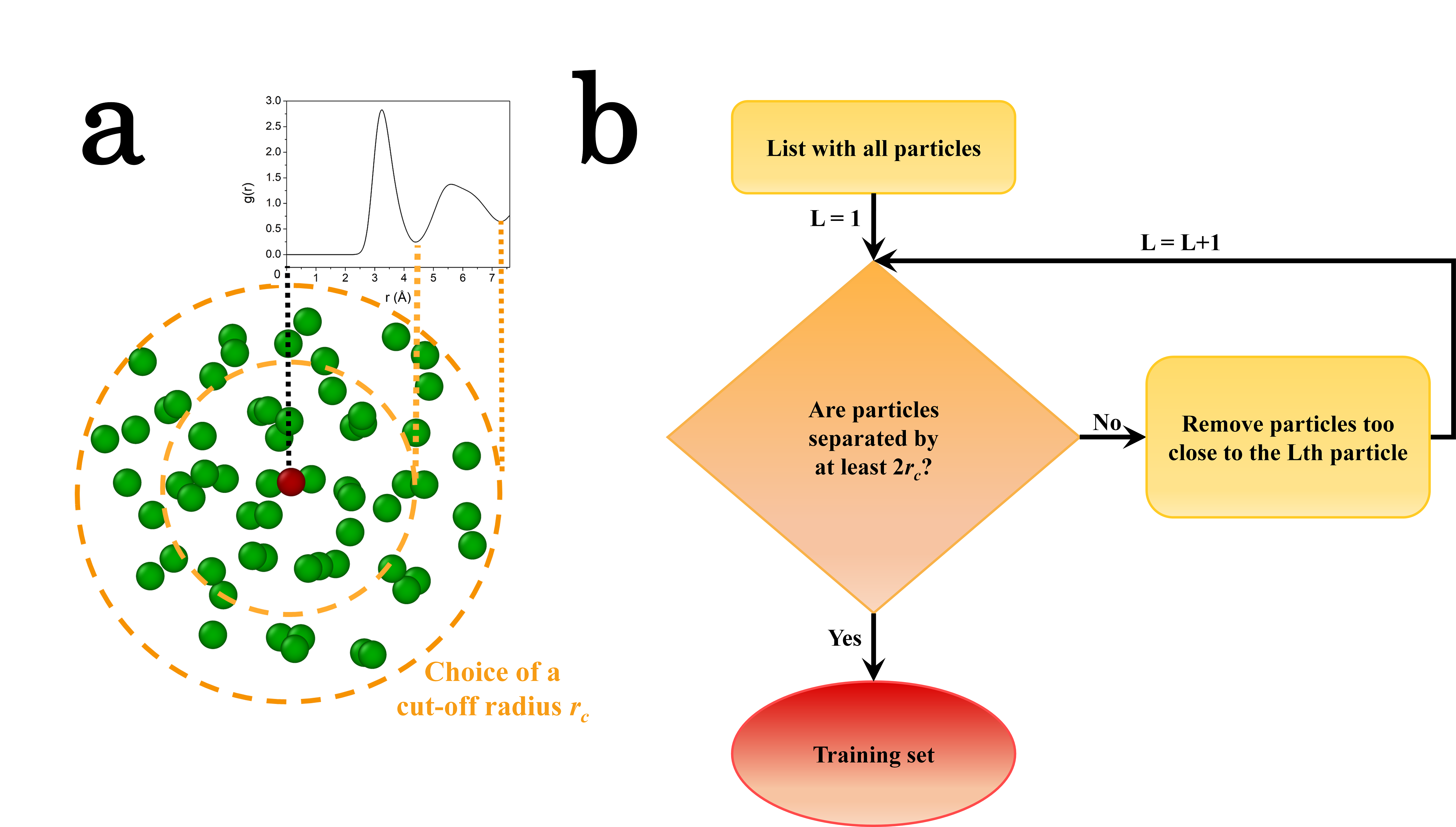}
	\caption{\label{fig:2}(a) Schematic illustration of the first and second neighbor shells of a central particle (in red) with cut-off radii determined by the radial distribution function g(r) of undercooled Zr at 1250 K. (b) Flowchart illustrating how to define a training set with independent structures. First, all particle positions in the simulation box are randomly ordered and set in a list. Then, the first particle of this list is set as a reference, and all particles which are closer than twice the selected cut-off radius $r_c$ from this reference particle are removed. Changing the reference particle, this process is iteratively repeated until all central particles meet the distance criterion based on $r_c$.}
\end{figure}

Local atomic structures in the configuration are defined through a coordination sphere, consisting of a central particle and its surrounding environment up to a chosen cut-off radius. The cut-off radius is defined thanks to the radial distribution function $g(r)$, which gives the probability to find a particle at a distance $r$ to a particle at the origin and is defined by
\begin{equation}
g(r)=\frac{N}{V}\frac{n(r)}{4\pi r^{2} \Delta r},
\end{equation}
where $n(r)$ stands for the mean number of particles in a spherical shell of radius $r$ and thickness $\Delta r$ centered on the origin. A typical representation of the radial distribution function $g(r)$ for the liquid state of Zr is depicted in Fig.~\ref{fig:2}(a). Each minimum of the radial distribution function can reasonably be used as a cut-off radius to obtain local environments that represent consecutive neighbor shells of each central particle. The first minimum (beyond the first maximum) defines the end of the first neighbor shell ($4.43$ Å from the central particle in Fig.~\ref{fig:2}(a)), the second minimum defines the end of the second neighbor shell ($7.28$ Å from the central particle) and so on.

Among the million particles in the configuration, local atomic structures are subsampled to construct a training set for the learning. To be representative of the configuration while satisfying statistical independence, the subsampling should pave the whole simulation box with respect to the PBC. For a given cut-off radius, the subsample is one among the (almost) maximal sets such that the central particles of the local atomic structures are separated by at least twice this cut-off radius. More precisely, such sets are computed through an iterative process summarized on the flowchart Fig.~\ref{fig:2}(b). The choice of the cut-off radius (between first shell, second shell, etc.) as well as the number of structures in the training set will be justified in Section \ref{Section:PH}.

Once the central particles are identified, the Python package \texttt{pyscal} \cite{Menon2019} is used on the full configuration to efficiently extract the particles coordinates of the local atomic structures. 

\subsection{Persistent homology as a local atomic structure descriptor}
\label{Section:PH}

To encode topological information of the local atomic structures, we use PH, a popular TDA tool \cite{Edelsbrunner2002, Zomorodian2005} that detects relevant topological features from a point cloud. 

\begin{figure}[h!]
	\centering
	\includegraphics[width=0.9\textwidth]{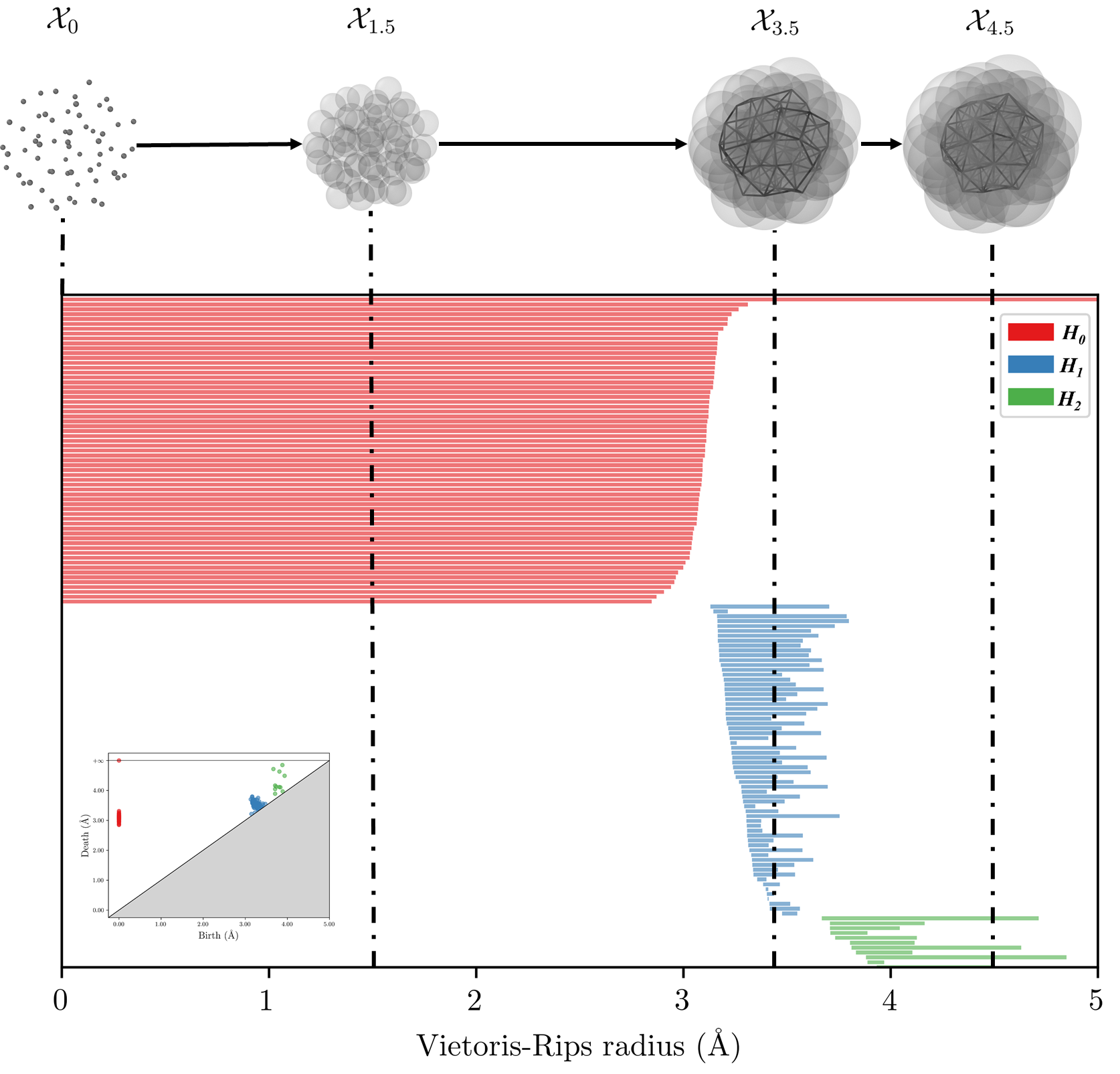}
	\caption{\label{fig:3}Schematic snapshots of the \textit{Vietoris-Rips complexes} filtration $(\mathcal X_t)_{t\geq 0}$ applied to a local structure with two neighbor shells. The \emph{barcode} below encodes the persistence of each topological feature by a line starting at its birth (the radius where its appear) and finishing at its death (the radius where it disappears). The colors correspond to the homological dimension (red, blue and green for $H_{0}$, $H_{1}$, and $H_{2}$ respectively) of each topological feature. The insert represents the corresponding PD where each point corresponds to a line, and thus its associated topological feature with coordinates (birth, death).}
\end{figure}

\begin{figure}[h!]
	\centering
	\includegraphics[width=1\textwidth]{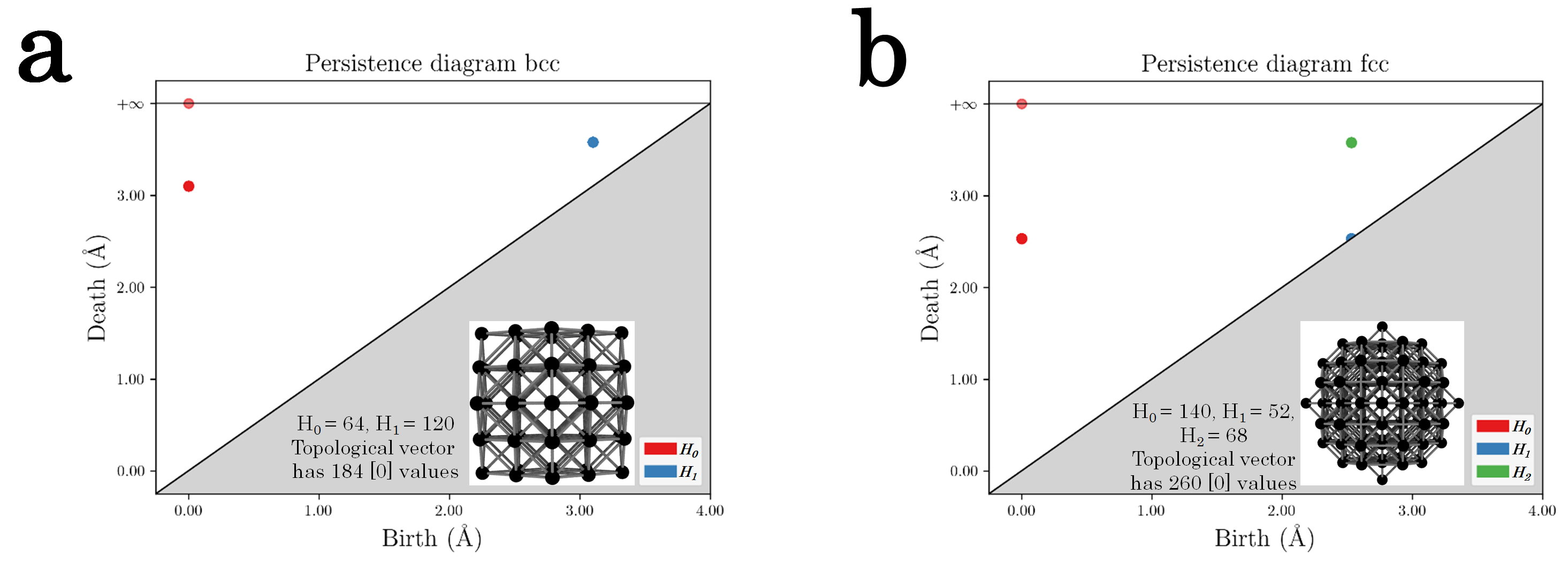}
	\caption{\label{fig:4}(a) Zr atoms on a bcc lattice and the corresponding PD. (b) Zr atoms on a fcc lattice and the corresponding PD. The homological dimensions $H_{0}$, $H_{1}$, and $H_{2}$ are depicted in red, blue, and green respectively. As these structures are on periodic lattices, all topological features of the filtration from a specific homological dimension have the same pair (birth, death).}
\end{figure}

Given a set of points $X$, one can construct a collection of simplicial complexes \cite{Ghrist2007}, called \textit{Vietoris-Rips complexes} as follows. Given $t\geq 0$, we consider the simplicial complex $\mathcal{X}_t$ which is the union of $k$-simplices for $k \in \mathbb{N}$, where 0-simplices are all the points of $X$, 1-simplices are segments with extremities in $X$ of length smaller than $t$, 2-simplices are full triangles with vertices in $X$ and distant of at most $t$ from one another, etc. In the end, we get an increasing sequence (called filtration) $\mathcal{X} = (\mathcal{X}_t)_{t\geq 0}$ of simplicial complexes. A schematic representation of this filtration process is depicted in Fig.~\ref{fig:3}. Actually, since the number of points is finite, changes in these complexes only appear at a finite sequence of steps $t_1<\cdots<t_n$ which gives a finite filtration of simplicial complexes $\mathcal{X}_0=\mathcal{X}_{t_0}\subset \mathcal{X}_{t_1}\subset \dots \subset\mathcal{X}_{t_n}$. For each space $\mathcal{X}_t$ of the filtration, we compute the simplicial homology, which is the sequence of their homology groups $(H_k(\mathcal{X}_t))_{k\geq 0}$, further denoted $(H_k)_{k \geq 0}$ when the space is understood or not specified for general consideration. The dimension of these homology groups gives insights on the dataset.  For instance, the dimension of $H_0$ gives the number of connected components, the dimension of $H_1$ the number of holes, and the dimension of $H_2$ the number of cavities inside the simplicial complex. Given a filtration of simplicial complexes $(\mathcal{X}_t)_{t\geq 0}$, one can keep track of topological features (encoded by elements in the homology groups) and how "persistent" they are, i.e., at which $t$ they appear (the birth) and at which $t$ they disappear (the death). A persistence diagram (PD) summarizes this information by plotting the pair (birth, death) of each topological feature of the filtration on a graph, as illustrated in Fig.~\ref{fig:3} or Fig.~\ref{fig:4}.

Using Python packages \texttt{gudhi} \cite{Maria2014} and \texttt{ripser.py} \cite{Tralie2018}, the individual PDs of the local atomic structures from the previously extracted training set are computed for each homological dimensions $H_{0}$, $H_{1}$, and $H_{2}$. To remove topological noise when considering $H_{1}$ and $H_{2}$, we use a subsampling approach as introduced in \cite{Fasy2014}.

While PDs are usually compared using the Bottleneck distance, their space equipped with this metric cannot be embedded into a Euclidean space nor even a normed vector space \cite{Carriere2019}. To tackle this problem, several mapping into vector spaces have been proposed in the literature. In this paper, we use a method developed in \cite{Carriere2015} and classically used to study 3D shapes. Each coordinate of the topological vector is associated to a pair of points $(x,y)$ in a PD $D$ for a fixed level of homology, except for the infinite point, and is calculated by 
\begin{equation}
m_D(x,y) = \min \{\|x-y\|_\infty, d_\Delta(x), d_\Delta(y)\},
\end{equation}
where $d_\Delta(\cdot)$ denotes the $\ell^\infty$ distance to the diagonal, and those coordinates are sorted by decreasing order. Remark that the dimension of each topological vector depends on the local atomic structure, so we fill in each vector with a non-informative value to reach the maximal dimension of the descriptor space. Here we decided to fill the vectors with the value $-1$ instead of $0$ as proposed in \cite{Carriere2015}, as distances between points in PDs are always nonnegative, and a zero, i.e., two points of the PD with the same birth and death, corresponds to relevant information in our context. 

To highlight the importance of these zeros in our topological vectors, let's consider Zr atoms on a body-centered cubic (bcc) lattice up to the second neighbor shell ($7.16$ Å from the central particle) and Zr atoms on a face-centered cubic (fcc) lattice in the same shell as shown in Fig.~\ref{fig:4}. Even though these structures are fundamentally different, it can be noted that the only thing that differs between the topological vectors obtained from each structure is the dimension of the vector (64+120 [0] values for the bcc lattice, 140+52+68 [0] values for the fcc lattice). Thus, if we fill the smaller with zeros, we will obtain two identical vectors, i.e., the same point in the descriptor space.

\begin{figure}[h!]
	\centering
	\includegraphics[width=1\textwidth]{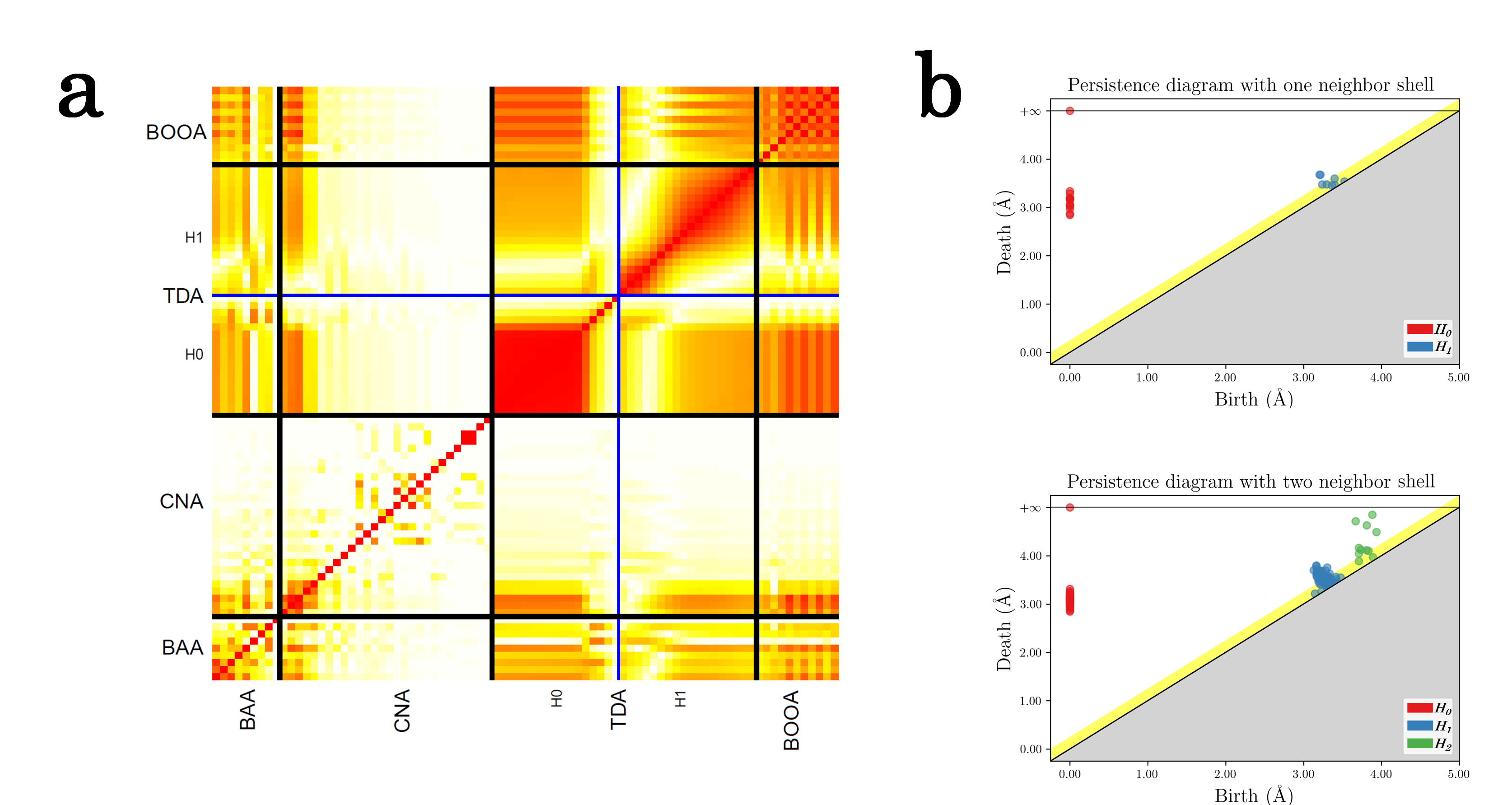}
	\caption{\label{fig:5}(a) Empirical correlation matrix (absolute value) between several descriptor families: BAA, CNA, TDA, and BOOA. Low values are white, high values are red. We distinguish between the descriptors by black lines, and between levels of homology with a blue line ($H_0$ and $H_1$). (b) PDs of a local atomic structure with one and two neighbor shells. The yellow line corresponds to a threshold computed with the subsampling approach to remove noises.}
\end{figure}

Finally, to illustrate the relevance of our topological signature to study the local atomic structures, we confront it on our Zr simulation to widely used classical physical descriptors such as the Bond Angle Analysis (BAA) \cite{Ackland2006}, the Common Neighbor Analysis (CNA) \cite{Honeycutt1987}, and the Bond-Orientational Order Analysis (BOOA) \cite{Steinhardt1983}, in its averaged definition \cite{Dellago2008}. The latter computes the order parameters $\bar{q}_{l}$ given by
\begin{equation}
\bar{q}_{l}(i)=\sqrt{\frac{4\pi}{2l+1}\sum_{m=-l}^{l}\lvert\frac{1}{\tilde{N}_{b}(i)}\sum_{k=0}^{\tilde{N}_{b}(i)}q_{lm}(k)\rvert^{2}}
\end{equation}
with the complex vector 
\begin{equation}
q_{lm}(i)=\frac{1}{N_{b}(i)}\sum_{j=1}^{N_{b}(i)}Y_{lm}(\mathbf{r}_{ij}),
\end{equation}
in which $\tilde{N}_{b}(i)$ is accounting for all the neighbors of a central particle $i$ plus $i$ itself, $N_{b}(i)$ for the sole number of nearest neighbors and $Y_{lm}(\mathbf{r}_{ij})$ are the spherical harmonics functions of the vector $\mathbf{r}_{ij}$ from particle $i$ to $j$, with $l$ an integer ranging from 2 to 12 and $m\in[-l,+l]$. A direct comparison of the CNA and BOOA, both being mostly used, can be found in Ref.~\cite{SupplementalMaterial3}. These classical descriptors characterize the radial and/or angular distribution of bonded pairs of atoms in the coordination sphere. The averaged version of the BOOA is preferred over the regular definition as a considerable improvement of the accuracy by taking into account information from the neighboring particles beyond the first neighbor shell. Fig.~\ref{fig:5}(a) gives an empirical correlation matrix between the coordinates of our TDA descriptor, BAA, BOOA, and CNA. This matrix is constructed on a sample of approximately $25$ $000$ local structures extracted as described in Section \ref{Section:DataPreparation} with a cut-off radius corresponding to the first minimum of the radial distribution function. The latter choice was made as a descriptor, as the CNA cannot be applied beyond the first shell of neighbors. The empirical correlation matrix shows that the topological signature is highly correlated with all the other descriptors. Some specificities of the data are highlighted: as all local neighborhood contain at least $10$ particles, the first $10$ components of $H_{0}$ are highly correlated; they are correlated with high values of $H_{1}$ too because most of the local structures have few $H_{1}$ components ($50$ \% of the population has less than 5 components); $H_{2}$ components are not depicted here as there is none in local atomic structures with only the first neighbor shell; the correlation matrix associated to the CNA is sparse, because only a few bonds, which are the first ones in the matrix (using the indexing of Faken and Jónsson \cite{Faken1994}: $[6 6 6]$ and $[4 4 4]$ for the bcc ordering and $[5 5 5]$, $[5 4 4]$ and $[4 3 3]$ for the icosahedral ordering) are present in most of the structures. It also explains the low correlation between the last bonds in the matrix of CNA with the other descriptors. 

This example on the first neighbor shell illustrates how relevant signatures from $H_0$ and $H_1$ are. To increase the number of $H_0$ and $H_1$ components and also to capture information on $H_2$ (which appears when considering more than just one neighbor shell), we built a training set of $5$ $314$ local atomic structures with two neighbor shells to construct our model, approaching the intermediate order. This leads here to $68$ $H_0$-components, $100$ $H_1$-components, and $22$ $H_2$-components, instead of only $17$ $H_0$-components and $18$ $H_1$-components with one neighbor shell. Fig.~\ref{fig:5}(b) shows a representative example of the differences in the PDs of a structure with one neighbor shell and the same structure with two neighbor shells. The second neighbor shell plays also a crucial role in the structural description \cite{Soper2000,Yan2007}.

Albeit considering higher-order shells is possible with our persistent homological description of  local structures, the advantage we gain in topological information is balanced by a too high spatial resolution of the local structures, leading to a loss of information in the GMM clustering. Thus, we have restricted ourselves to the second neighbor shell, through a trade-off between the structural information and a coarsening of the spatial resolution, already observed for the averaged BOOA \cite{Dellago2008}. Note that the reason for using the radial distribution function $g(r)$ to determine the second shell of neighbors is to fulfill this compromise; adaptive methods which are more general, such as SANN \cite{Meel2012} or RAD \cite{Higham2016} algorithms being limited to the first shell of neighbors. It is worth mentioning that the clustering remains essentially unchanged by varying the cut-off radius by $\pm5$ \% about the second minimum, which is compatible with the volume change during the crystal nucleation.

\subsection{Gaussian mixture model clustering}
\label{Section:LearningGMM}

A Gaussian Mixture Model (GMM) groups data points into clusters within an unsupervised way through a mixture of $M$ Gaussian distributions $(\phi(\;\cdot\;;\boldsymbol{\mu}_{m},\Sigma_{m}))_{1\leq m \leq M}$ of weights $(\alpha_{m})_{1\leq m \leq M}$ as
\begin{equation}
\sum_{m=1}^{M}\alpha_{m}\phi(\;\cdot\;;\boldsymbol{\mu}_{m},\Sigma_{m}),
\end{equation}
where $\boldsymbol{\mu}_{m}$ is the mean and $\Sigma_{m}$ the covariance matrix of the $m$th Gaussian distribution. After standardizing the descriptor space, an Expectation-Maximization (EM) algorithm \cite{Dempster1977} is used as an iterative method to estimate the unknown parameters $(\alpha_m, \mu_m, \Sigma_m)_{1\leq m \leq M}$. We use the implementation in the Python package \texttt{scikit-learn} \cite{Pedregosa2011} with full covariance matrices and $3$ $000$ $k$-means \cite{Lloyd1982} initializations.

To select the number of Gaussian components, i.e. the number of clusters, the integrated completed likelihood (ICL) \cite{Biernacki2000} criterion has been used: \begin{align}
	\text{ICL}&=-2\ln(\hat{L})+D\ln(n)-2\sum_{k=1}^{n}\sum_{m=1}^{M} \hat{\tau}_{m,k} ln( \hat{\tau}_{m,k} ) = \text{BIC}-2\sum_{k=1}^{n}\sum_{m=1}^{M}  \hat{\tau}_{m,k}  ln( \hat{\tau}_{m,k} ),
\end{align}
where $\hat L$ denotes the likelihood evaluated on the estimator, $D$ the number of parameters to be estimated ($D = M(d+d^2 + 1)$, with $d$ the dimension of the topological vectors), and $ \hat{\tau}_{m,k} $ denotes the probability to belong to the $m$th component conditionally to the observation $x_k$.
The ICL generalizes  the widely used Bayesian information criterion (BIC) \cite{Schwarz1978} to clustering methods by adding an entropic penalty computed from the posterior probabilities of the data to be assigned to each Gaussian component. The number of clusters in the model is  set to the number achieving the minimum of ICL.

\section{Application of the unsupervised approach on elemental zirconium simulations}
\label{Section:ApplicationApproach}
We focus in this section on the application of our method, called TDA-GMM, to elemental Zr simulations. It demonstrates that this protocol provides relevant structural information in a context as challenging as homogeneous nucleation.

\subsection{Learning the model}
\label{Section:LearningModel}

\begin{figure}[h!]
	\centering
	\includegraphics[width=0.5\textwidth]{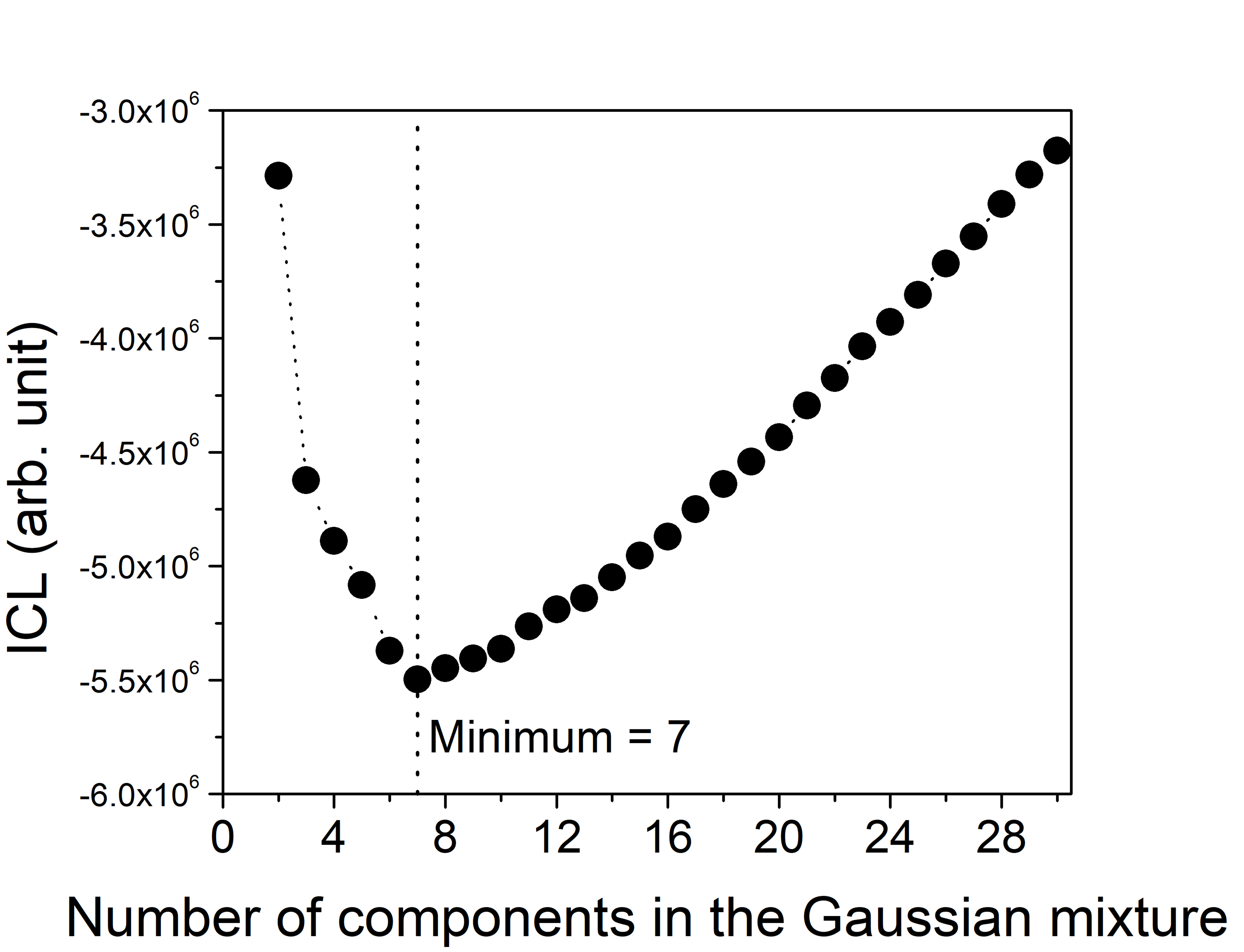}
	\caption{\label{fig:6}ICL criterion on the training set. Its minimum corresponds to the maximum likelihood, which leads to the optimal number of clusters to build the model.}
\end{figure}

\begin{figure}[h!]
	\centering
	\includegraphics[width=1\textwidth]{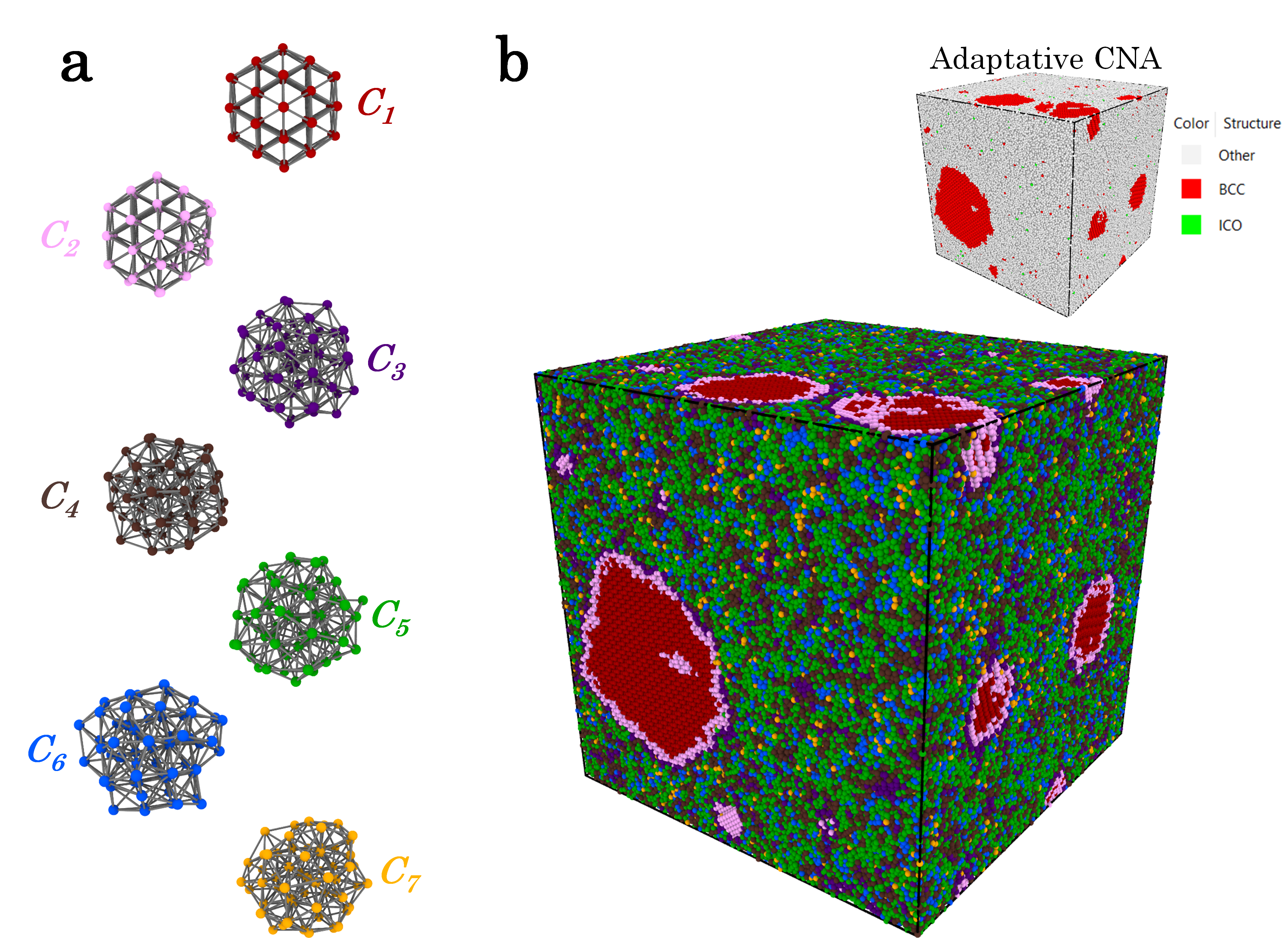}
	\caption{\label{fig:7}(a) Representation in the real space of the local structures assigned to each cluster $C_{i}$. (b) Snapshot of the one million atoms in the simulation box after the clustering in the descriptor space. In the inset, the classification obtained from an adaptative CNA method is depicted.}
\end{figure}

As mentioned in Sec. \ref{Section:DataPreparation},  configuration is chosen in course of the nucleation process, where the supercooled liquid coexists with crystalline nuclei, to capture all structural atomic events of interest. Fig.~\ref{fig:6} depicts the ICL computed on the extracted training set as described in Section \ref{Section:PH}, leading to select $7$ clusters \cite{SupplementalMaterial}. Each cluster $C_{i}$ is represented by the local structure that is the closest to the mean of the corresponding Gaussian component, as illustrated in Fig.~\ref{fig:7}(a). Note that here, an embedded data representation of the space of interest is not straightforward to interrogate visually the width and shape of the different Gaussians \cite{SupplementalMaterial2}, but an analysis of the eigenvalues of the covariance matrices shows different elliptical shapes. This proves the necessity of GMM over simpler unsupervised algorithms such as $k$-means which would only be suitable to fit hyperspheres.

Once the model is learned, it can be applied to new uncategorized local atomic structures through their topological signatures. The robustness of our model is checked by considering a test set consisting of all the structures in the configuration but the one in the training set. Most of the structures in this test set (99.998 \%) have probabilities beyond 0.999 to be in one cluster of the model, and the remaining ones still have probabilities higher than 0.5. Further analysis of the fit of the model  to the test set is performed using the  Mahalanobis distance (MaD) \cite{Mahalanobis1936}, which computes the distance between an empirical Gaussian distribution and a  dataset. Using hard-assignment to affect each structure to a cluster, we compute the MaD for each cluster between the test set and the multivariate Gaussian distribution. Outliers are then considered based on the interquartile range rule. Table~\ref{Maha} presents the percentage of structures in each cluster viewed as outliers according to this criterion: more than 96 \% of the structures in the test set have a MaD in agreement with their assigned Gaussian distribution of the model.

\begin{table}[h!]
	\begin{tabular}{crrrrrrrr}
		\hline
		\hline
		Clusters & \makebox[1.5cm][r]{$C_{1}$} & \makebox[1.5cm][r]{$C_{2}$} & \makebox[1.5cm][r]{$C_{3}$} & \makebox[1.5cm][r]{$C_{4}$} & \makebox[1.5cm][r]{$C_{5}$} & \makebox[1.5cm][r]{$C_{6}$} & \makebox[1.5cm][r]{$C_{7}$} & \makebox[1.5cm][r]{Total}\\
		\hline
		Proportion (\%) & 11.70 & 7.22 & 9.28 & 24.45 & 33.46 & 11.03 & 2.86 & 100\\
		\hline
		Outliers (\%) & 2.47 & 2.12 & 6.29 & 3.25 & 3.67 & 3.88 & 3.06 & 3.56\\
		\hline
	\end{tabular}	
	\caption{Proportions of structures and percentage of outliers in each cluster of the test set based on the interquartile range rule on the MaD between the test set and their assigned multivariate Gaussian distribution of the model in the descriptor space.}
	\label{Maha}
\end{table}

The final clustering is shown in Fig.~\ref{fig:7}(b) on the entire configuration depicted in the real space of the simulation box with the help of  \textsc{ovito}  \cite{Stukowski2010}. In the inset, the classification found by a classical adaptative CNA algorithm as implemented in \textsc{ovito} is represented for comparison. In contrast to TDA-GMM, only two types of structures that correspond to strictly crystalline or icosahedral structures were successfully identified.

\subsection{Description of the bulk crystal and liquid}
\label{Section:LiquidCrystal}

\begin{figure}[h!]
	\centering
	\includegraphics[width=0.8\textwidth]{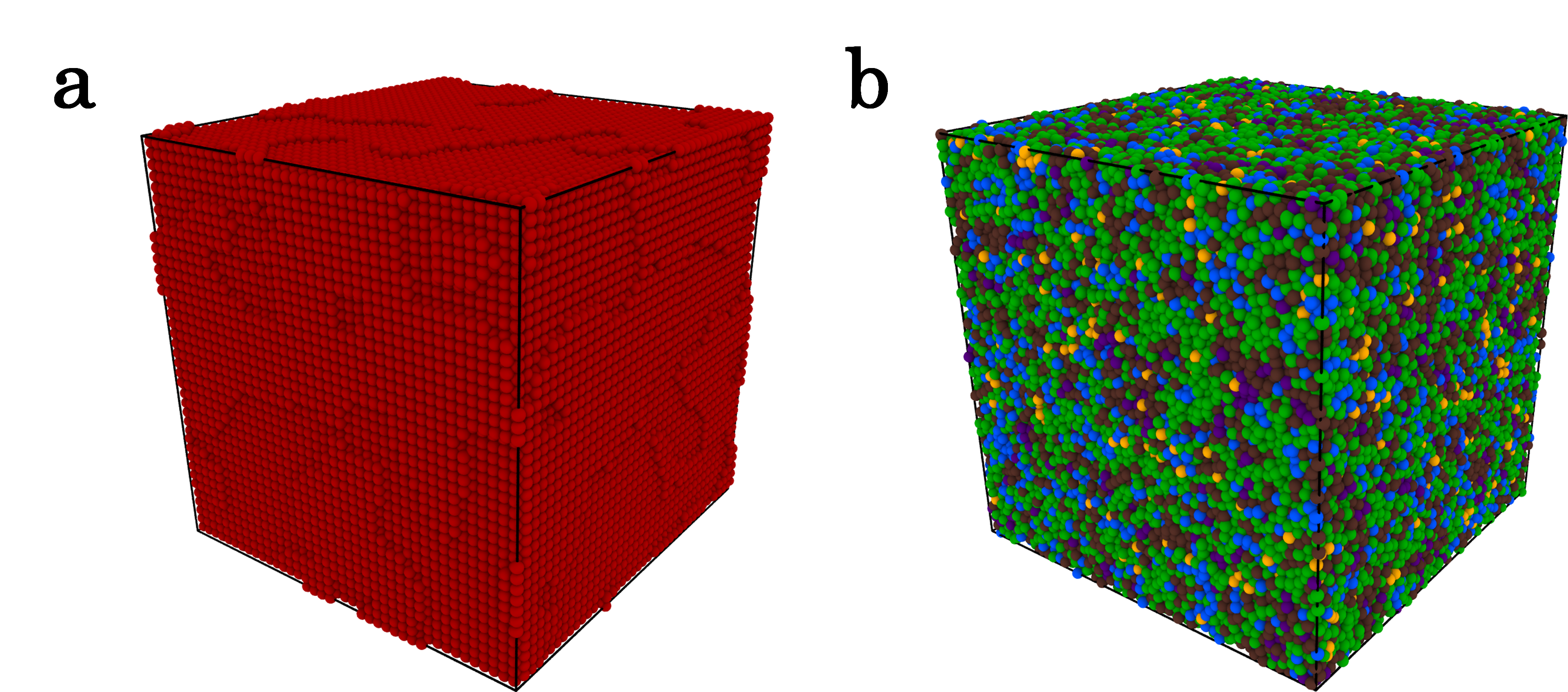}
	\caption{\label{fig:8}Snapshots at $T=1250$ K of Zr bulk crystal (a) and liquid (b) brought in their inherent local structures. Atoms are colored according to the cluster they belong to (see Fig.~\ref{fig:7}).}
\end{figure}

We have first applied our model to a configuration with $N=128$ $000$ atoms in the bulk crystal and liquid at $T=1250$ K. Fig.~\ref{fig:8} and Table~\ref{Tab:bulk} shows the results obtained in both cases through the central atoms of the local structures associated with each cluster.

\begin{table}[h!]
	\begin{tabular}{crr}
		\hline
		\hline
		Bulk & \makebox[2cm][r]{Crystal} & \makebox[2cm][r]{Liquid}\\
		\hline
		$C_{1}$ (\%) & 100 & 0.00 \\
		$C_{2}$ (\%) & 0 & 0.03 \\
		$C_{3}$ (\%) & 0 & 6.49 \\
		$C_{4}$ (\%) & 0 & 29.83 \\
		$C_{5}$ (\%) & 0 & 45.22 \\
		$C_{6}$ (\%) & 0 & 14.70 \\
		$C_{7}$ (\%) & 0 & 3.72 \\
		\hline
	\end{tabular}	
	\caption{Proportion of each cluster $C_{i}$ in the bulk crystal and liquid.}
	\label{Tab:bulk}
\end{table}

The structures of the bulk crystal are classified as local structures in cluster $C_{1}$. Referring to Fig.~\ref{fig:7}(a), the atoms of the mean local structure associated with $C_{1}$ are indeed on a periodic lattice. In the bulk liquid, more than 99.96 \% of the identified local structures belong to clusters $C_{3}, C_4, C_5, C_6$ and $C_{7}$. From this simple application, and without any information  about the physical nature of the atomic structures, the trained TDA-GMM model distinguishes solid-like from liquid-like clusters.

\subsection{Description of the homogeneous crystal nucleation}
\label{Section:Nucleation}

During the process of homogeneous nucleation  along an isotherm, the internal energy of the system undergoes a sharp drop. It follows the growth of the nuclei that carry the crystalline periodicity, until the state of least energy is reached. To autonomously detect these nuclei among the clusters obtained by the TDA-GMM method, five configurations along an isotherm at the nose of the TTT at $T=1250$ K with $N=1$ $024$ $000$ atoms are selected. They range from the onset of nucleation at $160$ ps from the quench, up to a configuration in the poly-crystalline bulk at $1560$ ps. The model built in Section \ref{Section:LearningModel} corresponds to a configuration in the course of nucleation at $360$ ps. Table~\ref{Tab:ST-ClustersZr} shows the evolution of the proportion of each cluster in these configurations.

\begin{table}[h!]
	\begin{tabular}{crrrrrr}
		\hline
		\hline
		Time (ps) & \makebox[2cm][r]{160} & \makebox[2cm][r]{200} & \makebox[2cm][r]{240} & \makebox[2cm][r]{280} & \makebox[2cm][r]{310$^{(M)}$} & \makebox[2cm][r]{1560$^{(S)}$}\\
		\hline
		$C_{1}$ (\%) & 0.25 & 0.95 & 2.85 & 6.83 & 12.66 & 67.21 \\
		$C_{2}$ (\%) & 0.67 & 1.37 & 2.76 & 4.96 & 6.85 & 22.65 \\
		$C_{3}$ (\%) & 7.37 & 7.68 & 8.18 & 8.75 & 8.51 & 6.99 \\
		$C_{4}$ (\%) & 29.99 & 29.47 & 28.41 & 26.52 & 24.22 & 1.58 \\
		$C_{5}$ (\%) & 43.69 & 42.87 & 40.99 & 37.42 & 33.78 & 0.90 \\
		$C_{6}$ (\%) & 14.44 & 14.14 & 13.48 & 12.36 & 11.16 & 0.35 \\
		$C_{7}$ (\%) & 3.59 & 3.53 & 3.32 & 3.15 & 2.82 & 0.31 \\
		\hline
	\end{tabular}	
	\caption{Proportion of each cluster $C_{i}$ at different times of the nucleation process. Superscripts (M) and (S) correspond respectively to the configuration used to train the model and the solidified configuration.}
	\label{Tab:ST-ClustersZr}
\end{table}

\begin{figure}[h!]
	\centering
	\includegraphics[width=1\textwidth]{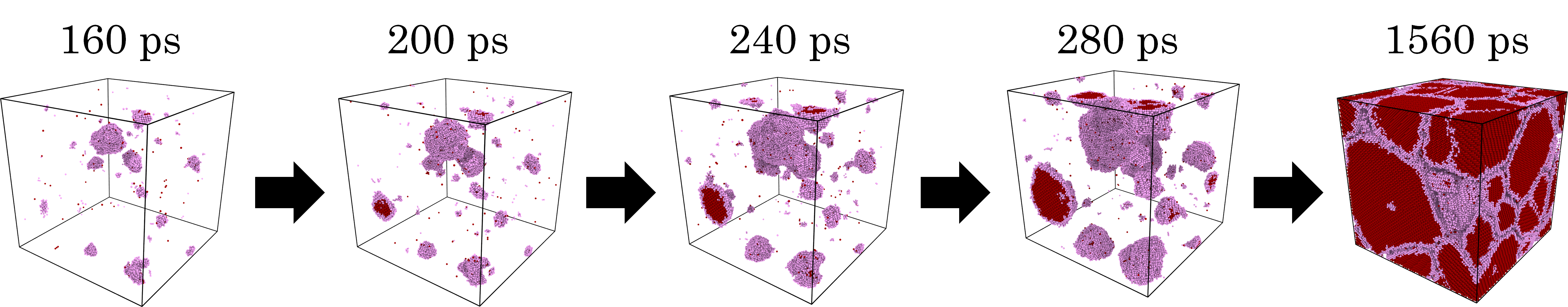}
	\caption{\label{fig:9}Snapshots of Zr's configurations along nucleation at the nose of the TTT at $T=1250$ K. The local structures in the clusters $C_{1}$ and $C_{2}$ identified as carrying the crystalline periodicity are depicted through their central particle in red and pink, respectively.}
\end{figure}

Local structures from clusters $C_{1}$ and $C_{2}$ follow a fast growth until they become the majority in the final bulk. On the opposite, the other clusters undergo a more or less strong decrease in their proportions over time. As mentioned in Section \ref{Section:LiquidCrystal}, local structures from $C_{1}$ correspond to a crystalline ordering whereas atoms of the mean local structure associated with $C_{2}$ are on a partial periodic lattice. Fig.~\ref{fig:9} represents the snapshots of these clusters in these different configurations. The other clusters, classified as liquid, show that apart from the nuclei, there exist complex structural heterogeneities in the supercooled liquid, which we describe in Section \ref{Section:LiquidHeterogeneities}, and which are successfully highlighted by the TDA-GMM method. Such an assessment has recently been brought out as well in supercooled liquid mixtures by another unsupervised method based on an information-theoretic approach \cite{Paret2020}.

The independent nuclei formed by local structures from $C_{1}$ and/or $C_{2}$ are extracted using a distance-based neighboring criterion as implemented in \textsc{ovito}. This criterion is set as the first minimum of $g(r)$ in such a manner that two particles are connected if they are at a distance less than or equal to this threshold. The size distributions of these independent nuclei are then obtained by accounting for all atoms of the overlapping local atomic structures of two neighbor shells defined by each central particle belonging to $C_{1}$ and/or $C_{2}$. The critical size is estimated by the smallest cluster that persists over the several configurations of nucleation, with at least the same number of atoms. In such a high undercooling regime, this leads  to a critical nucleus size of 70-80 atoms.

\subsection{Translational and orientational orderings}
\label{Section:OrderingAnalysis}

Two physical orderings which have been shown to be important parameters driving nucleation are the translational and orientational orderings \cite{Russo2012, Russo2016, Berryman2016}.

With the nuclei previously identified, their centers of mass are easily inferred. Based on these positions, the spatial evolution of the translational ordering from the nuclei to the liquid can be quantified through density. The density is indeed the dedicated indicator to measure the fluctuations in the relative positions of neighboring atoms: for each cluster $C_{i}$ and its relative number of atoms $K_i(r_{n})$ in a spherical shell of volume $V_{s}$ extending from the center of a nucleus to a distance $r_{n}$ to $(r_{n}+\epsilon)$ with $\epsilon=1$ \AA, the radial partial atomic density profiles $\rho_i(r_{n})$ are computed as:
\begin{align} 
	\rho_i(r_{n}) &= \frac{K_i(r_{n})}{V_{s}}, \hspace{1cm}
	\text{with } V_{s} = \frac{4}{3}\pi [(r_{n}+\epsilon)^{3}-r_{n}^{3}].
\end{align}	

On the other hand, the analysis of the orientational ordering provides information on the fluctuations in relative geometric bonding between neighboring atoms. Such a characterization can be performed with help of the classical methods mentioned in Section \ref{Section:PH}. For instance, the BOOA has shown to be successful to analyze the crystallization of colloidal suspensions \cite{Gasser2001, Gasser2003}. Here, following the indexing of Faken and Jónsson, the CNA was performed in each cluster as a representative measure \cite{Jakse2003}. It was verified, as indicated by the high correlations in Fig.~\ref{fig:5}(a), that these two methods give similar information about the orientational ordering \cite{SupplementalMaterial3}.

\begin{figure}[h!]
	\centering
	\includegraphics[width=0.65\textwidth]{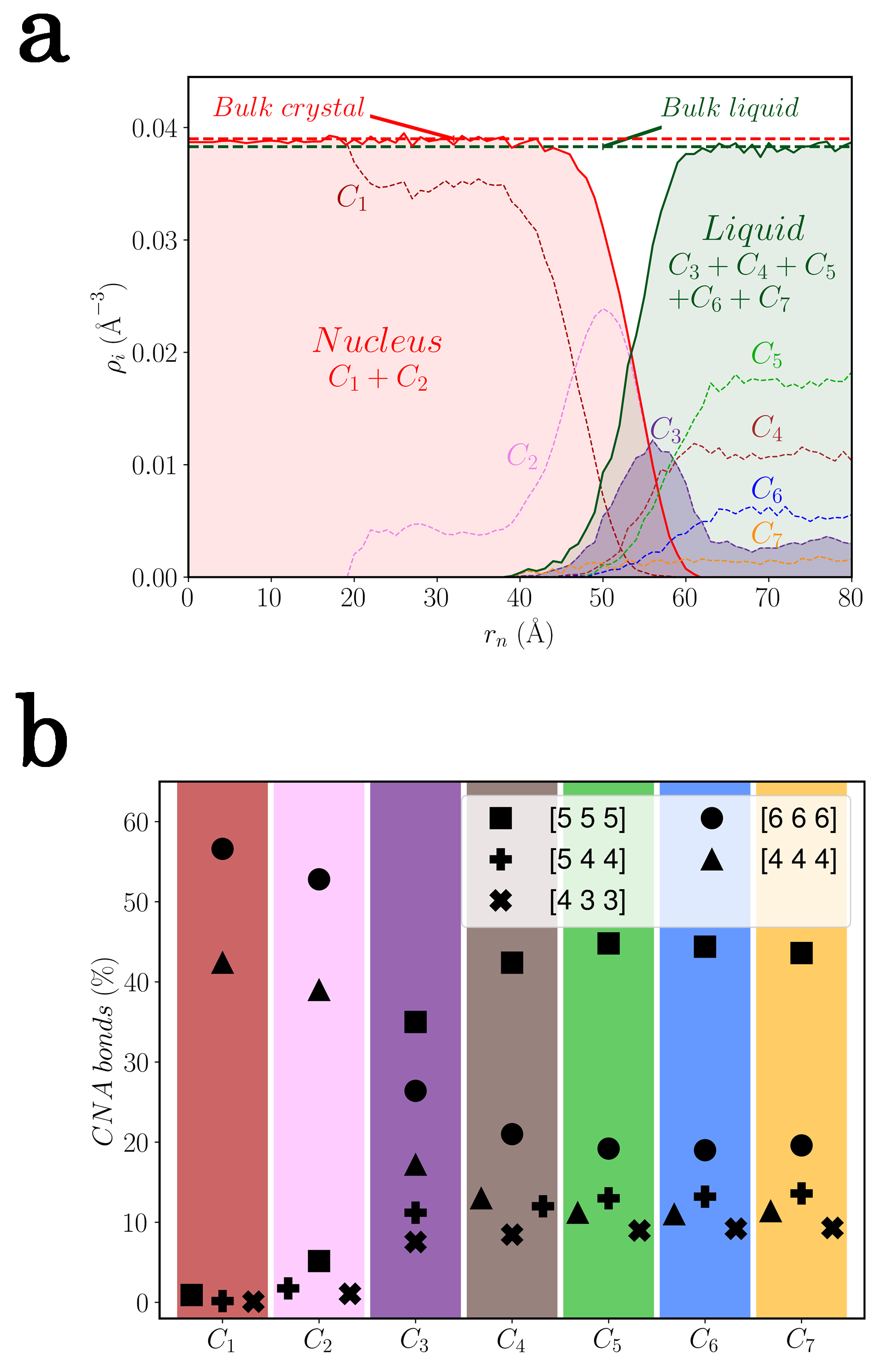}
	\caption{\label{fig:10}(a) Representation on a nucleus of the typical translational ordering of the nuclei and their surrounding liquid through the radial density profile of the $C_{i}$ clusters. The corresponding densities of the bulk crystal and undercooled liquid have been determined at $T=1250$ K and constant ambient pressure. This behavior of the translational ordering is representative of all nuclei in the configurations, even the precritical ones. (b) Typical orientational ordering of the $C_{i}$ clusters in the configurations through CNA. $[555]$, $[544]$ and $[433]$ stand for five-fold symmetries bonds and, $[666]$ and $[444]$ for bonds relative to a bcc crystal ordering.}
\end{figure}

This both orderings analysis shows us a general behavior, even for the precritical nuclei, which is depicted in Fig.~\ref{fig:10}. Fig.~\ref{fig:10}(a) reveals that all the nuclei emerge with a translational ordering which corresponds to the density of the bulk crystal. Despite the small density window between the bulk crystal and liquid at $T=1250$ K, it is clearly spotted that the sum of the respective densities $\rho_{i}$ of the clusters $C_{i}$ associated with the nuclei or liquid leads to the density values of the bulk crystal and liquid, respectively. Fig.~\ref{fig:10}(b) confirms the crystalline structures in $C_{1}$ and $C_{2}$ which exhibit bcc bonds ($[666]$: 8, $[444]$: 6) and thus a concurrent emergence of the orientational ordering in the nucleation process. The slight percentage of icosahedral bonds $[555]$ in $C_{2}$ tends to explain the partial periodicity observed in Fig.~\ref{fig:7}(a). On the other hand, the local structures from clusters $C_{4}$ to $C_{7}$ share strong five-fold symmetries bonds characteristics of the undercooled liquid, although they also possess non-negligible relative bcc bonds. About $4$ $\%$ of $[666]$ in these clusters are in excess to form bcc structures and have the potential to combine with $[555]$ bonds to form Z14, Z15, and Z16 Frank-Kasper polyhedra \cite{Nelson1989}. This can lead to a geometrical frustration able to slow down the onset of the nucleation in the liquid part \cite{Tanaka2012,Russo2018}. At last, $C_{3}$ is a more peculiar case, with a strong distribution at the border of nuclei while also being present in the deeper liquid. Its presence in the frontier region is explained by the spatial resolution already mentioned in Section \ref{Section:PH}, while its presence in the liquid can be seen as a precursor for the emergence of the embryos. This is in agreement with the latest view of a heterogeneous scenario in which precursors of the crystalline ordering arise from structural heterogeneities in the liquid \cite{Pasturel2017,Shibuta2017,Jug2021} and which we develop in Section \ref{Section:LiquidHeterogeneities}.

There is no consensus in the literature for a general pathway of the nucleation process of materials, which seems to be system dependent: hard spheres systems undergo concurrent translational and orientational ordering \cite{Berryman2016}, whereas a decoupling of these two symmetries has been observed in colloidal systems \cite{Russo2012}. For a pure metal like Zr, our unsupervised analysis brings us to the conclusion that the translational and orientational orderings arise synchronously.

\subsection{Liquid heterogeneities}
\label{Section:LiquidHeterogeneities}

To assess the scenario mentioned in Section \ref{Section:OrderingAnalysis}, heterogeneity at an early nucleation stage is evaluated by a comparison of the distribution of atoms belonging to each cluster against the uniform distribution. As the comparison of distributions is particularly complex in three dimensions since the empirical cumulative distribution function is not defined, a Kolmogorov-Smirnov (KS) test \cite{Kolmogoroff1933} was performed on the projection of atomic positions in the three directions of space. For a level $0.01$, if the uniform distribution is rejected on at least one projection, it is rejected for the entire multivariate dataset. Table~\ref{Tab:KS-Zr} depicts the results through the $p$-values of the test obtained from the configuration at $160$ ps. For all the clusters the test is always rejected, meaning that their distributions are not uniform, in other words, indicating structural heterogeneities in the simulation box.

\begin{table}[h!]
	\centering	
	\begin{tabular}{crrr}
		\hline
		\hline
		$p$-value & \makebox[2cm][r]{$x$} & \makebox[2cm][r]{$y$} & \makebox[2cm][r]{$z$}\\
		\hline
		$C_{1}$ & 0.000 & 0.000 & 0.000 \\
		$C_{2}$ & 0.000 & 0.000 & 0.000 \\
		$C_{3}$ & 0.000 & 0.000 & 0.000 \\
		$C_{4}$ & 0.000 & 0.000 & 0.007 \\
		$C_{5}$ & 0.001 & 0.000 & 0.000 \\
		$C_{6}$ & 0.007 & 0.331 & 0.001 \\
		$C_{7}$ & 0.192 & 0.156 & 0.005 \\
		\hline
		\hline
	\end{tabular}
	\caption{$p$-values computed on the projection of atomic positions at an early nucleation stage ($160$ ps) onto each direction of the simulation box from a KS test against the uniform distribution.}
	\label{Tab:KS-Zr}
\end{table}

\section{Conclusions}
\label{Section:Conclusion}

We used PH, a main computational tool in TDA, to build descriptors of local atomic structures. A GMM allowed us to autonomously identify clusters of similar local structures in the space of these topological descriptors. We applied this unsupervised approach to the analysis of several MD configurations of Zr in the challenging context of homogeneous nucleation. We highlighted some interesting properties of the phenomenon such as a concurrent emergence of the translational and orientational orderings and structural heterogeneities in the undercooled liquid. A closer look at the nucleation process of other pure metals has been investigated through this TDA-GMM method and the results will be published in future work. An extension of the analysis on multicomponent systems where chemical orderings arise would be especially relevant for further inspection and control of such mechanisms to enhance materials design. More generally, this unsupervised learning methodology opens the door to further study of structure-dependent phenomena at the atomic scale.

\section*{Acknowledgement}

We acknowledge the CINES and IDRIS under Project No. INP2227/72914, as well as CIMENT/GRICAD for computational resources. This work was performed within the framework of the Centre of Excellence of Multifunctional Architectured Materials “CEMAM”ANR-10-LABX-44-01 funded by the “Investments for the Future” Program. This work has been partially supported by MIAI@Grenoble Alpes (ANR-19-P3IA-0003). Fruitful discussions within the French collaborative network in high-temperature thermodynamics GDR CNRS3584 (TherMatHT) are also acknowledged.

\renewcommand{\thefigure}{S\arabic{figure}}
\renewcommand{\thetable}{S\Roman{table}}

\begin{center}
{\Large Supplemental Material}
\end{center}
\maketitle

\section*{Boxplots of the descriptors for each cluster obtained by the TDA-GMM method on elemental Zr}

\begin{figure}[H]
	\centering
	\includegraphics[width=1\textwidth]{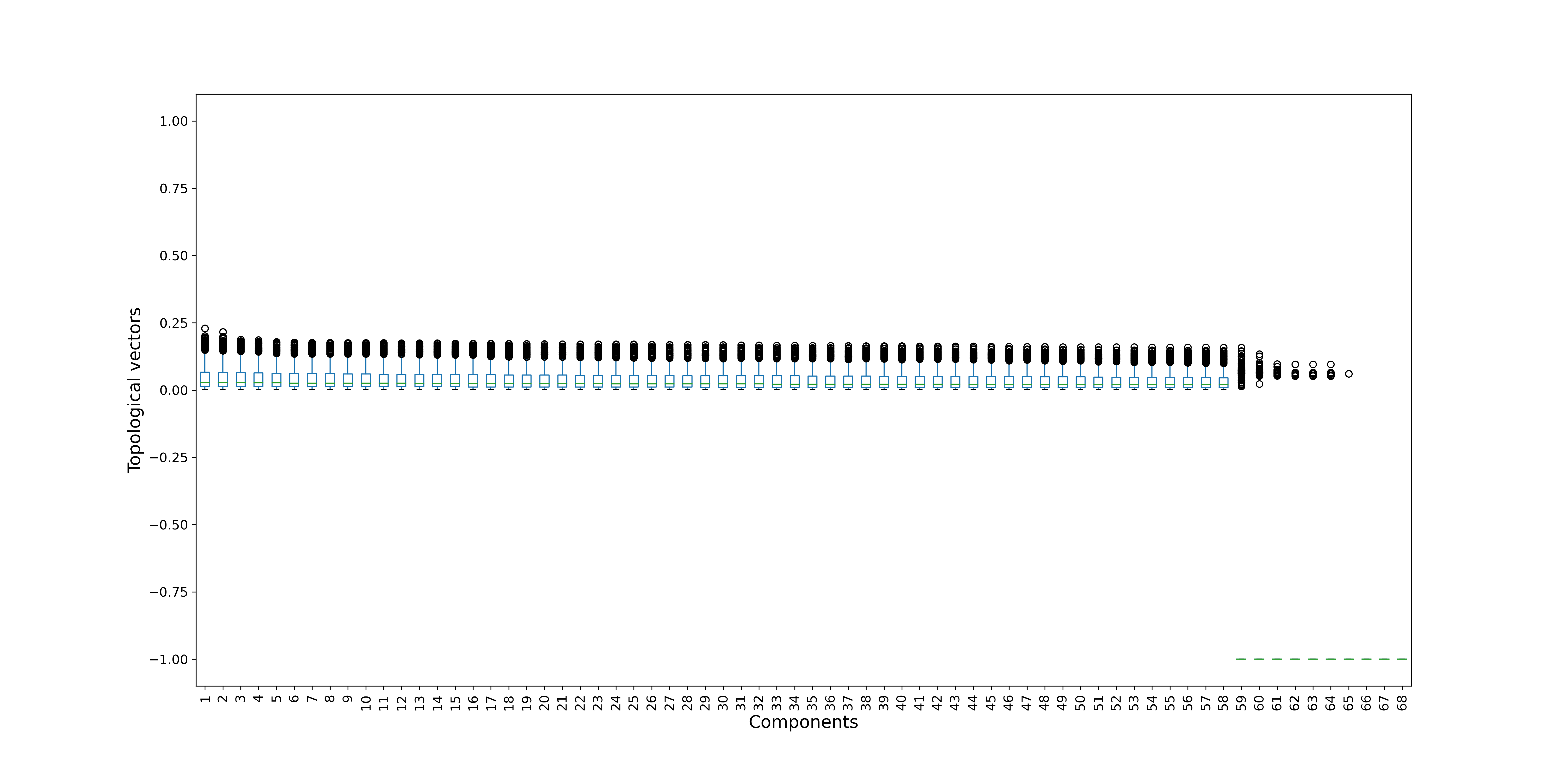}
	\caption{\label{c1_0}Boxplots of the topological vectors in homological dimension $H_{0}$ for data in $C_{1}$.}
\end{figure}

\begin{figure}[H]
	\centering
	\includegraphics[width=1\textwidth]{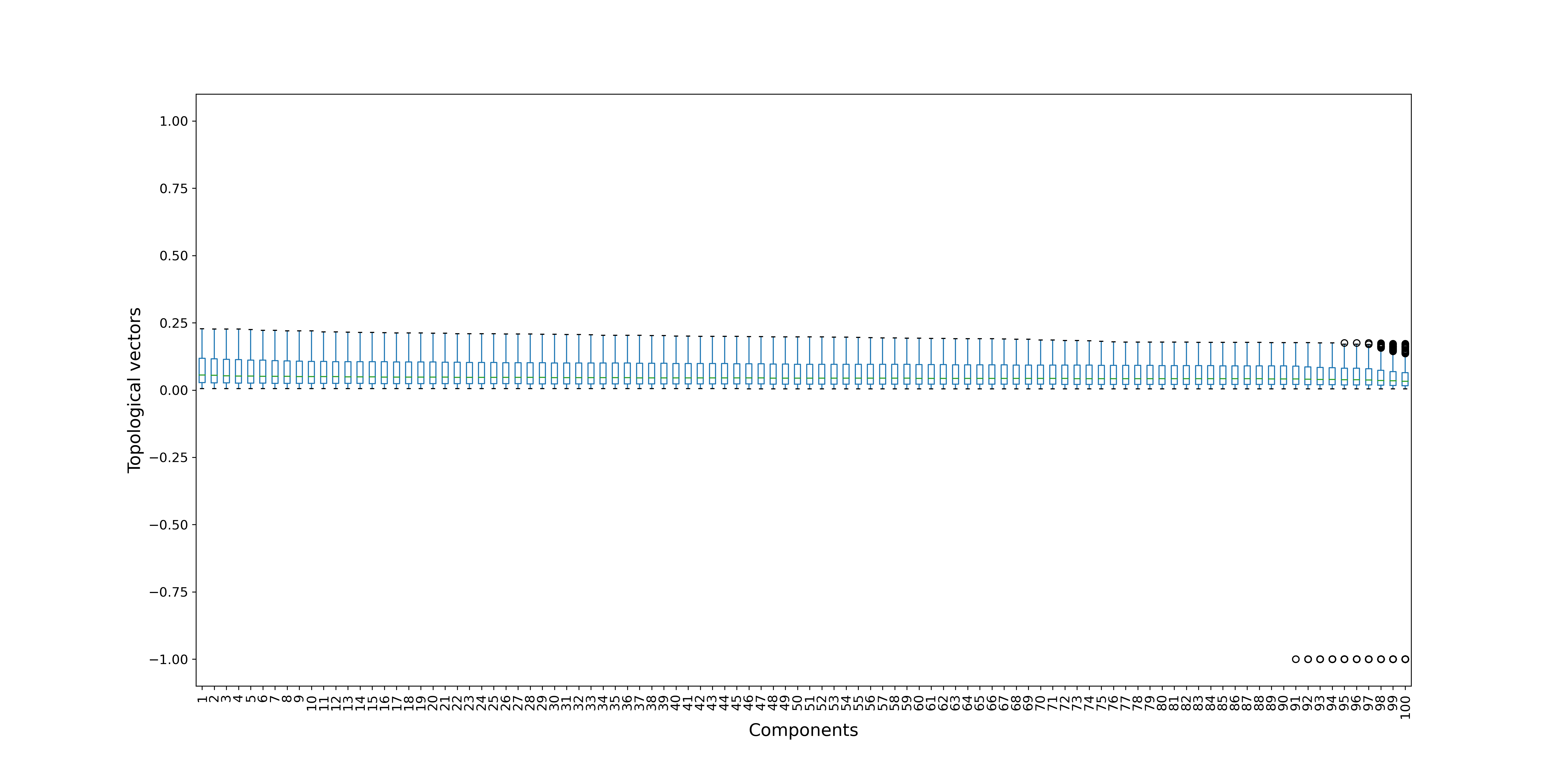}
	\caption{\label{c1_1}Boxplots of the topological vectors in homological dimension $H_{1}$ for data in $C_{1}$.}
\end{figure}

\begin{figure}[H]
	\centering
	\includegraphics[width=1\textwidth]{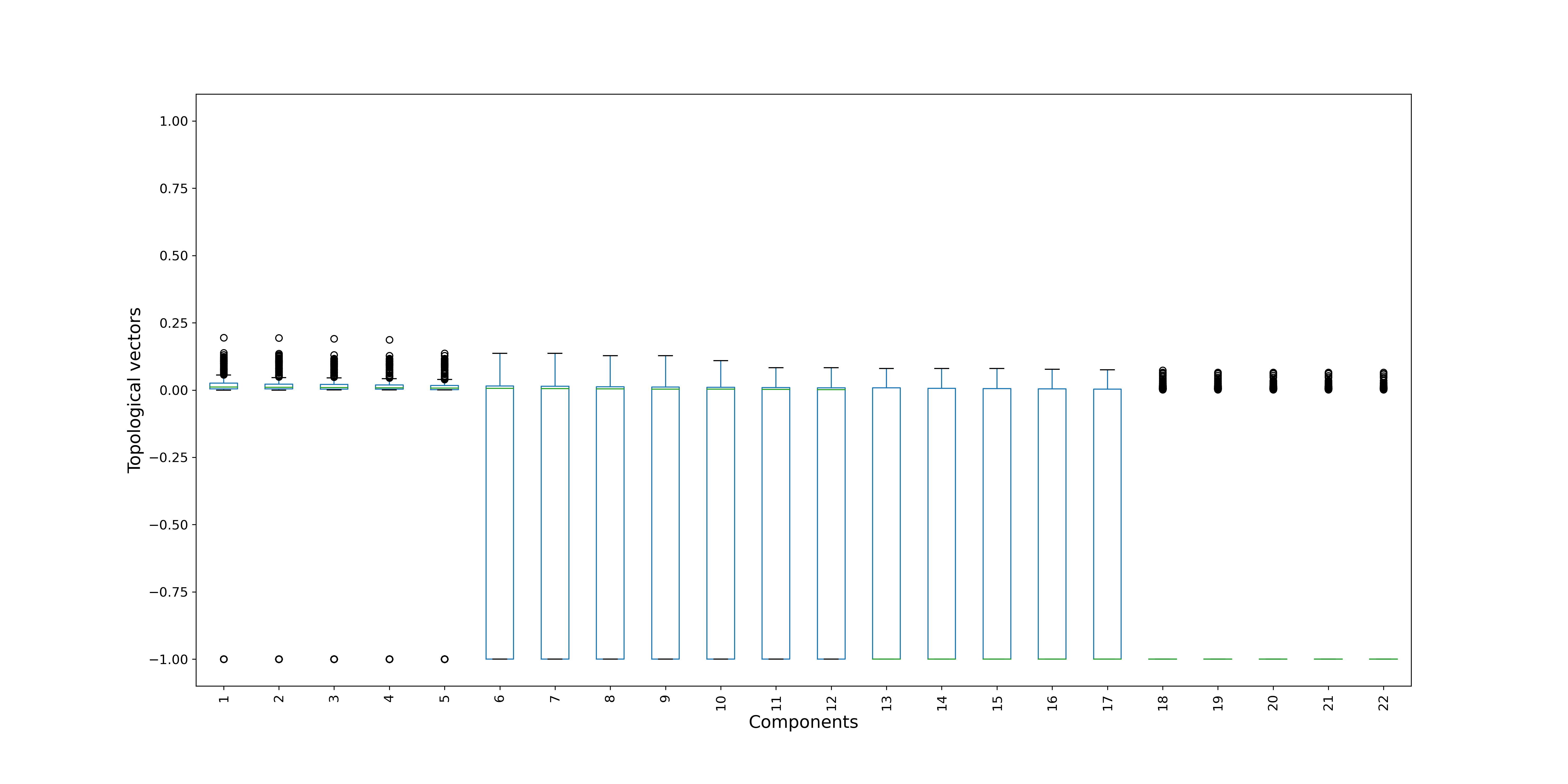}
	\caption{\label{c1_2}Boxplots of the topological vectors in homological dimension $H_{2}$ for data in $C_{1}$.}
\end{figure}

\begin{figure}[H]
	\centering
	\includegraphics[width=1\textwidth]{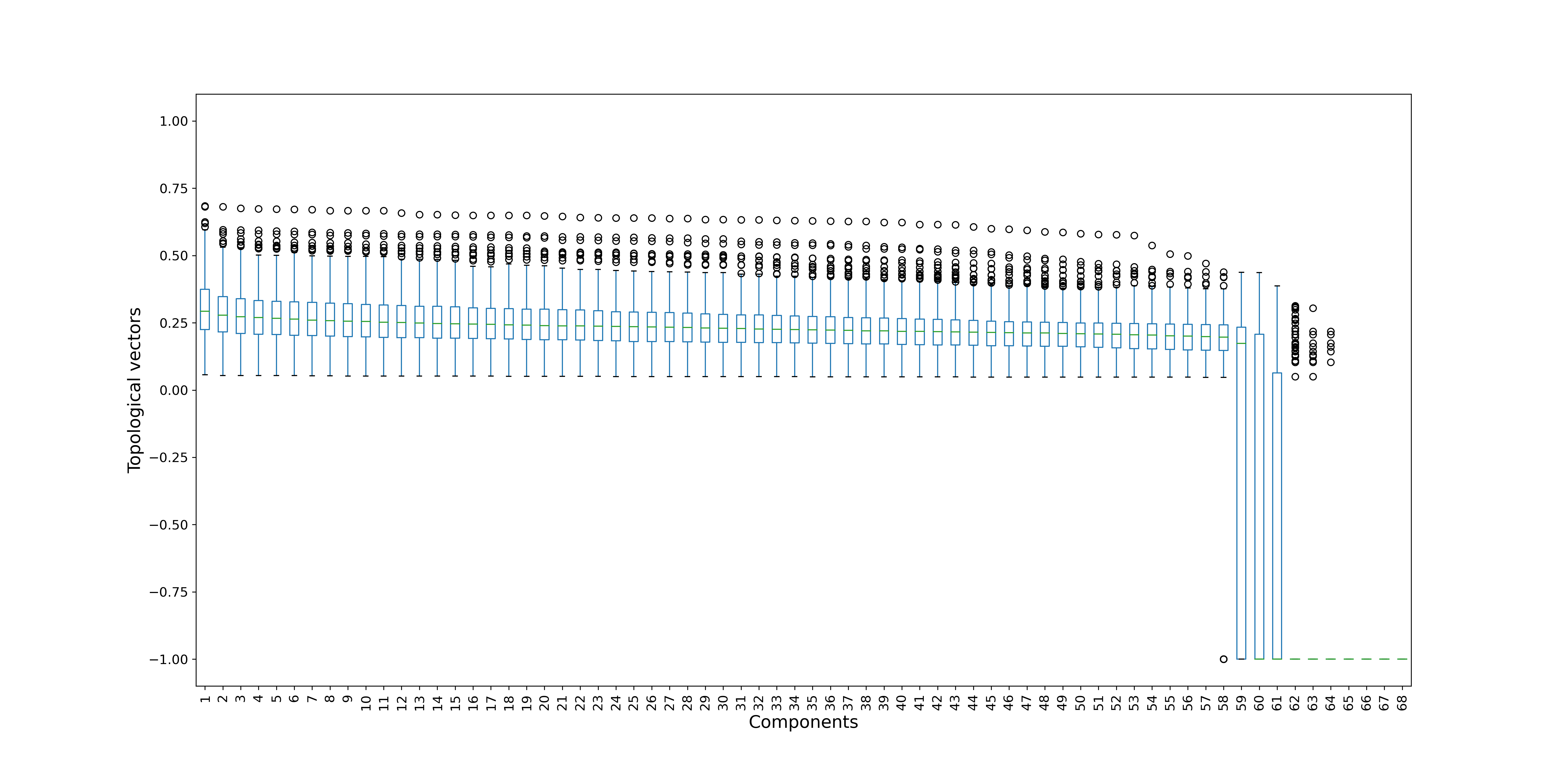}
	\caption{\label{c2_0}Boxplots of the topological vectors in homological dimension $H_{0}$ for data in $C_{2}$.}
\end{figure}

\begin{figure}[H]
	\centering
	\includegraphics[width=1\textwidth]{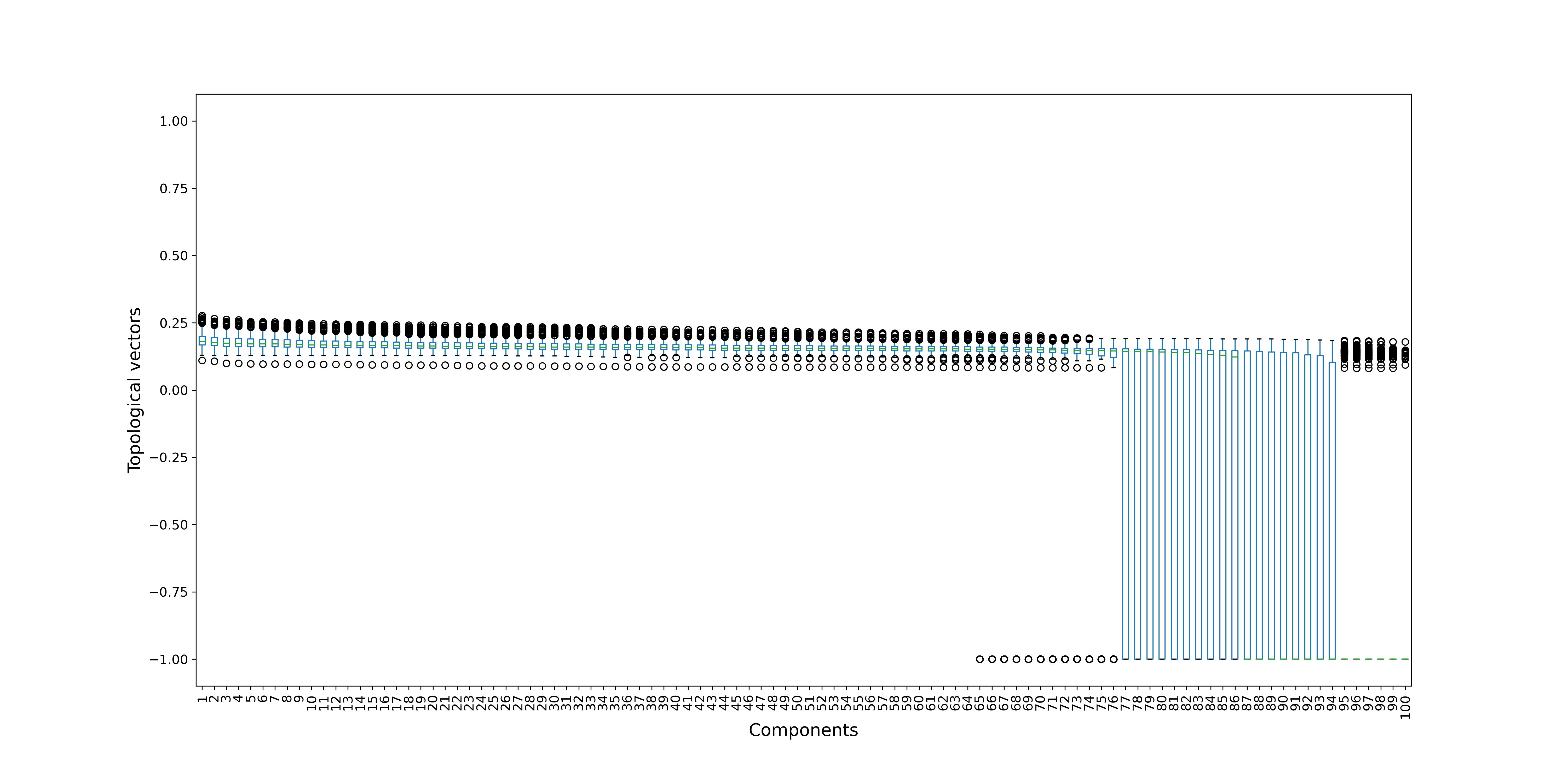}
	\caption{\label{c2_1}Boxplots of the topological vectors in homological dimension $H_{1}$ for data in $C_{2}$.}
\end{figure}

\begin{figure}[H]
	\centering
	\includegraphics[width=1\textwidth]{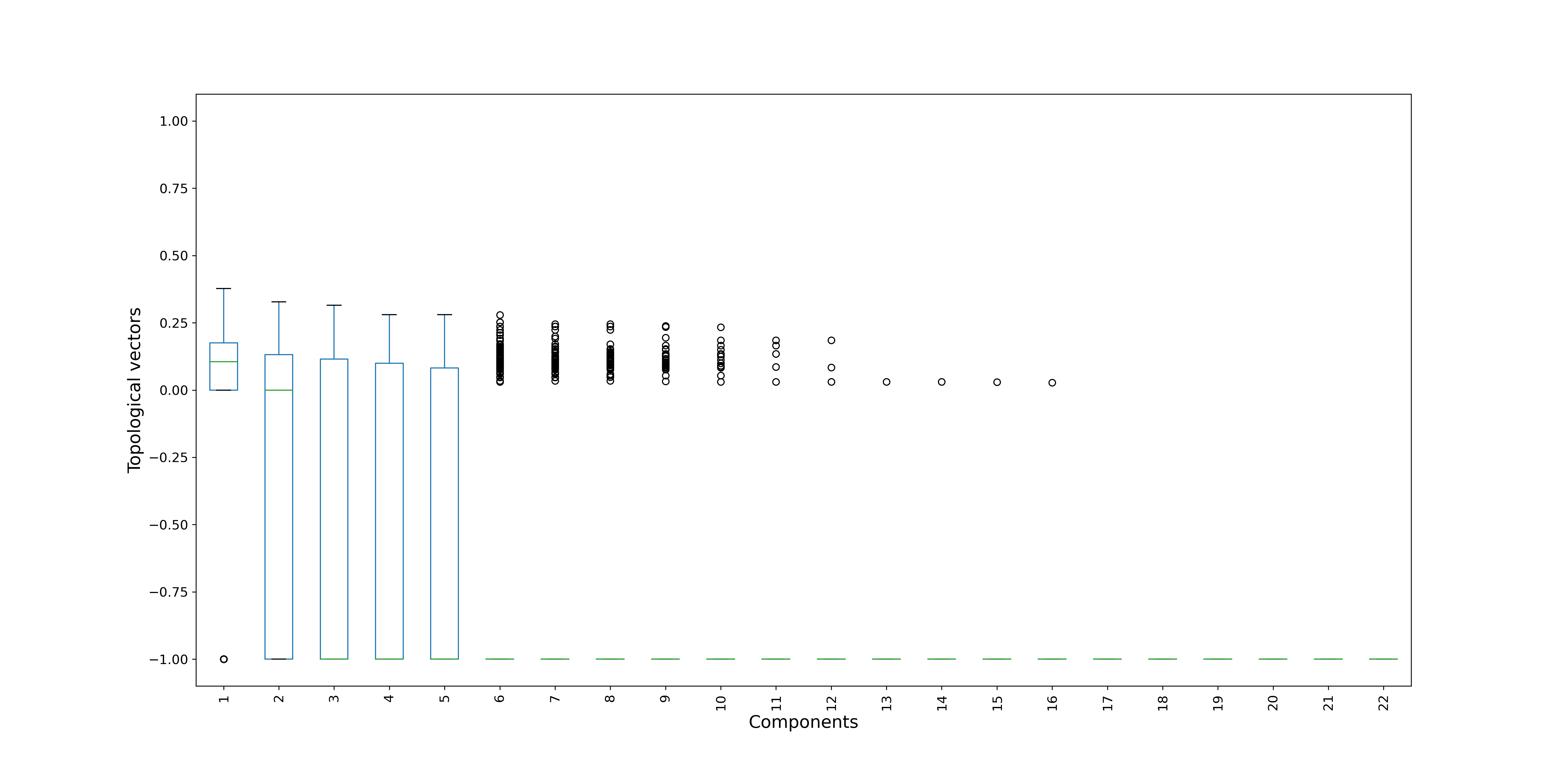}
	\caption{\label{c2_2}Boxplots of the topological vectors in homological dimension $H_{2}$ for data in $C_{2}$.}
\end{figure}

\begin{figure}[H]
	\centering
	\includegraphics[width=1\textwidth]{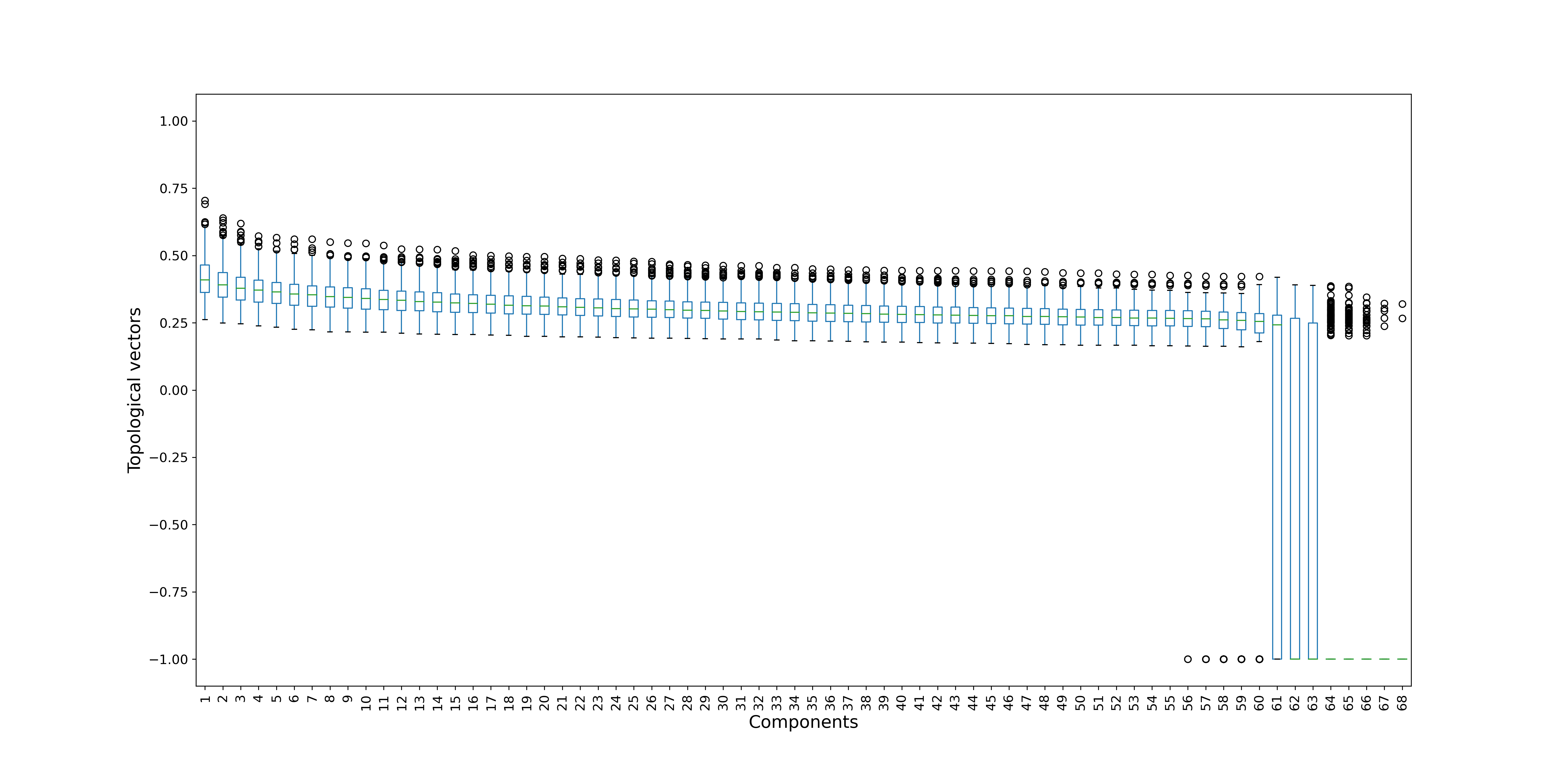}
	\caption{\label{c3_0}Boxplots of the topological vectors in homological dimension $H_{0}$ for data in $C_{3}$.}
\end{figure}

\begin{figure}[H]
	\centering
	\includegraphics[width=1\textwidth]{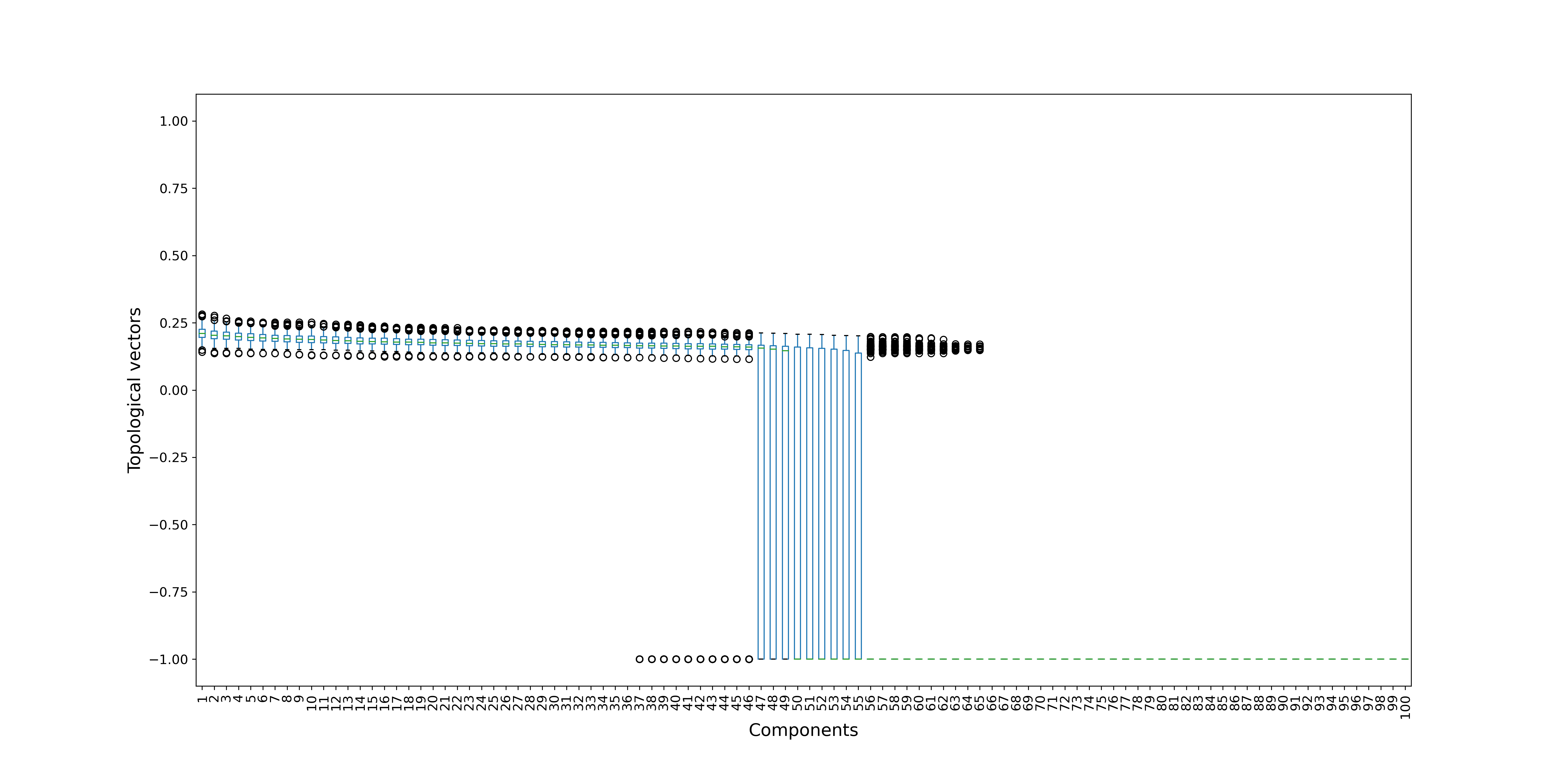}
	\caption{\label{c3_1}Boxplots of the topological vectors in homological dimension $H_{1}$ for data in $C_{3}$.}
\end{figure}

\begin{figure}[H]
	\centering
	\includegraphics[width=1\textwidth]{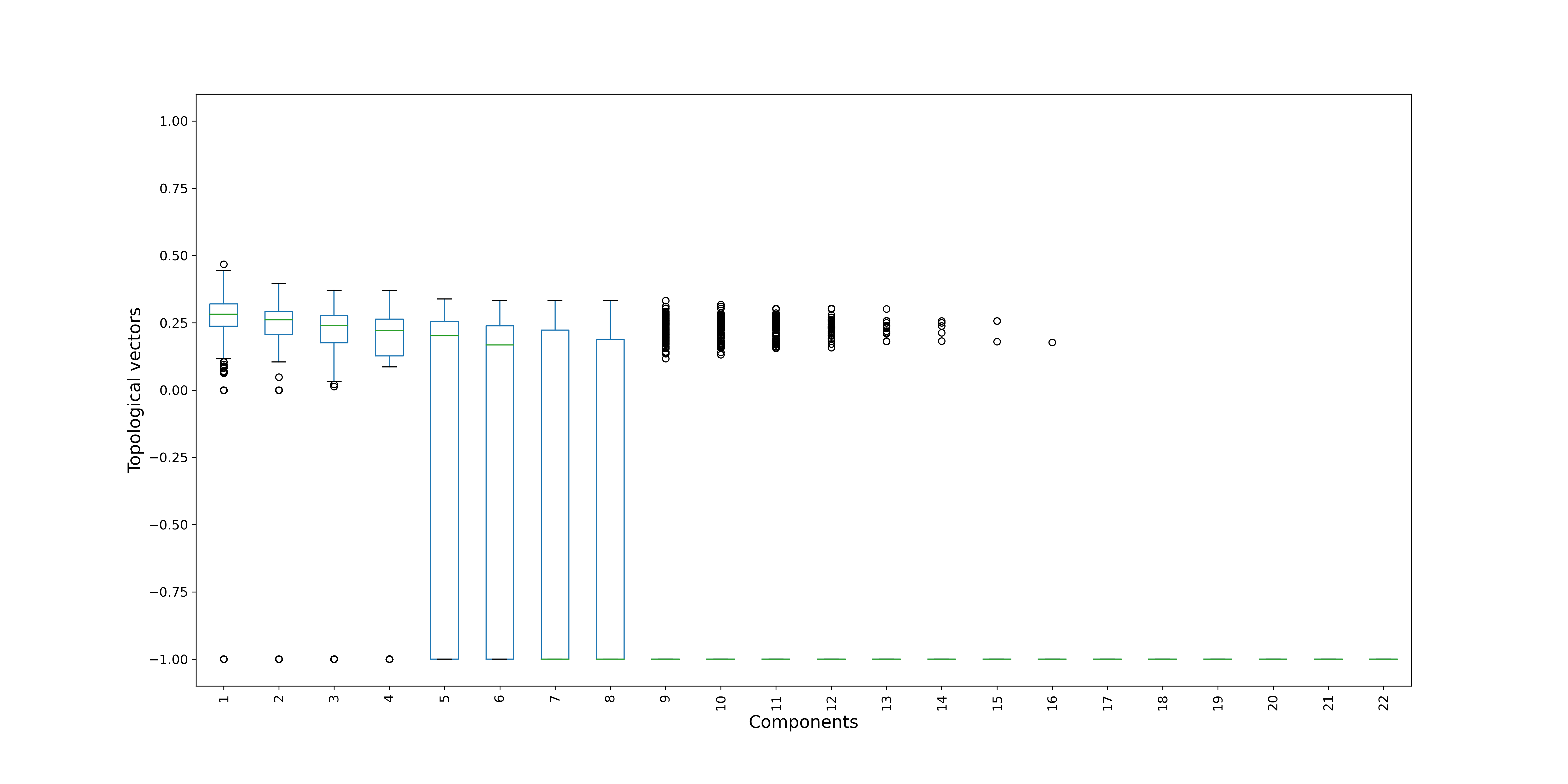}
	\caption{\label{c3_2}Boxplots of the topological vectors in homological dimension $H_{2}$ for data in $C_{3}$.}
\end{figure}

\begin{figure}[H]
	\centering
	\includegraphics[width=1\textwidth]{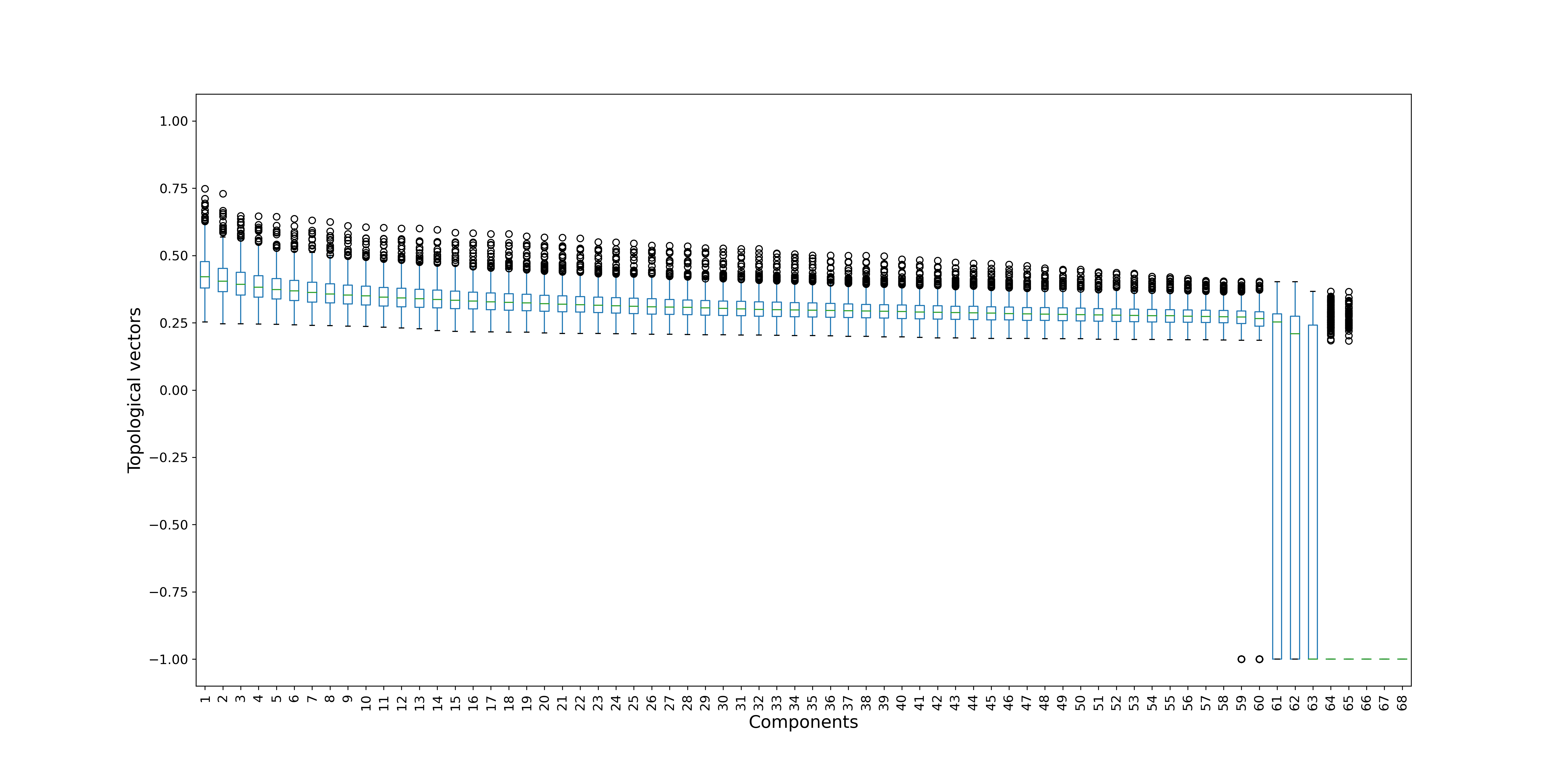}
	\caption{\label{c4_0}Boxplots of the topological vectors in homological dimension $H_{0}$ for data in $C_{4}$.}
\end{figure}

\begin{figure}[H]
	\centering
	\includegraphics[width=1\textwidth]{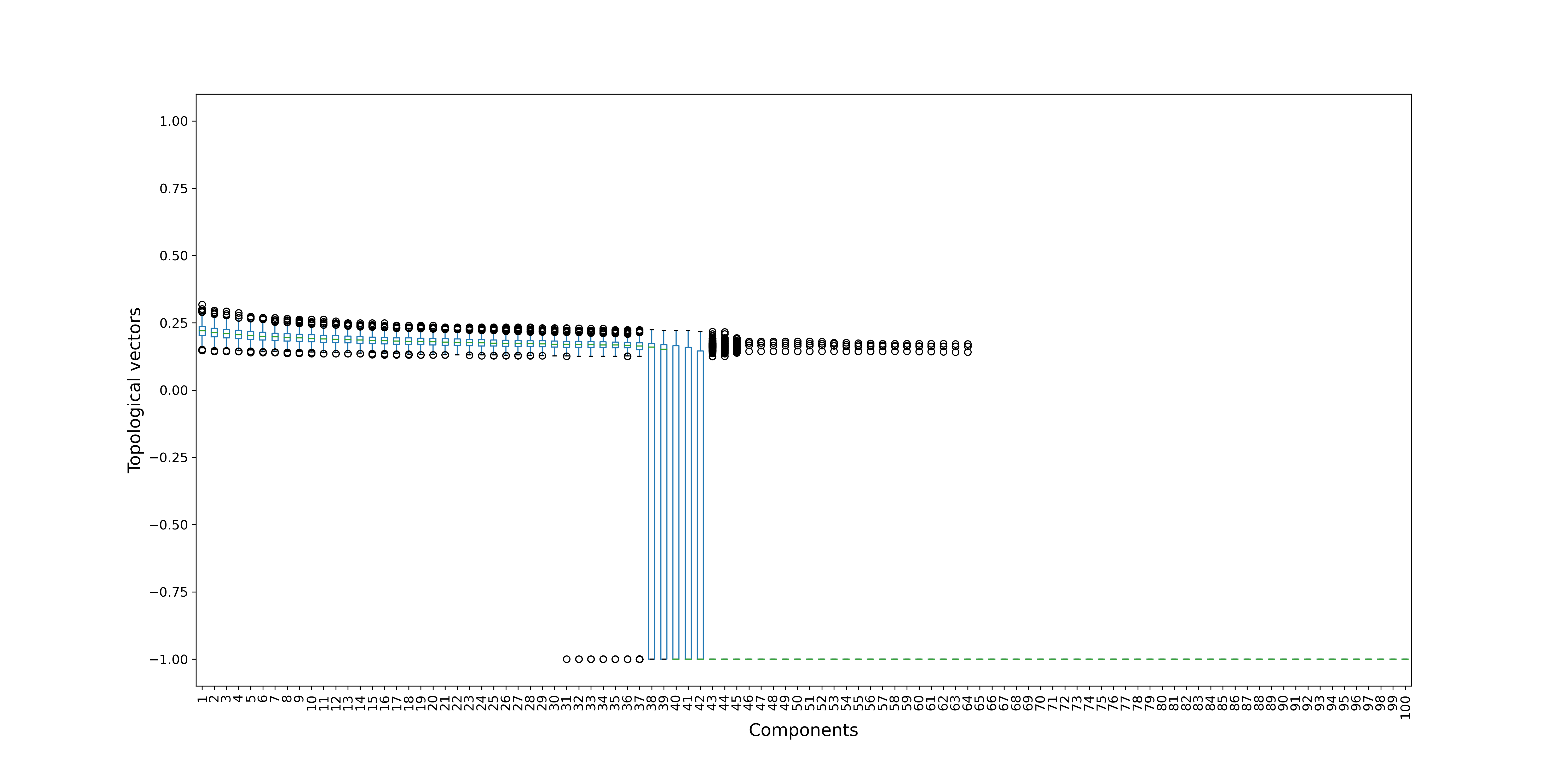}
	\caption{\label{c4_1}Boxplots of the topological vectors in homological dimension $H_{1}$ for data in $C_{4}$.}
\end{figure}

\begin{figure}[H]
	\centering
	\includegraphics[width=1\textwidth]{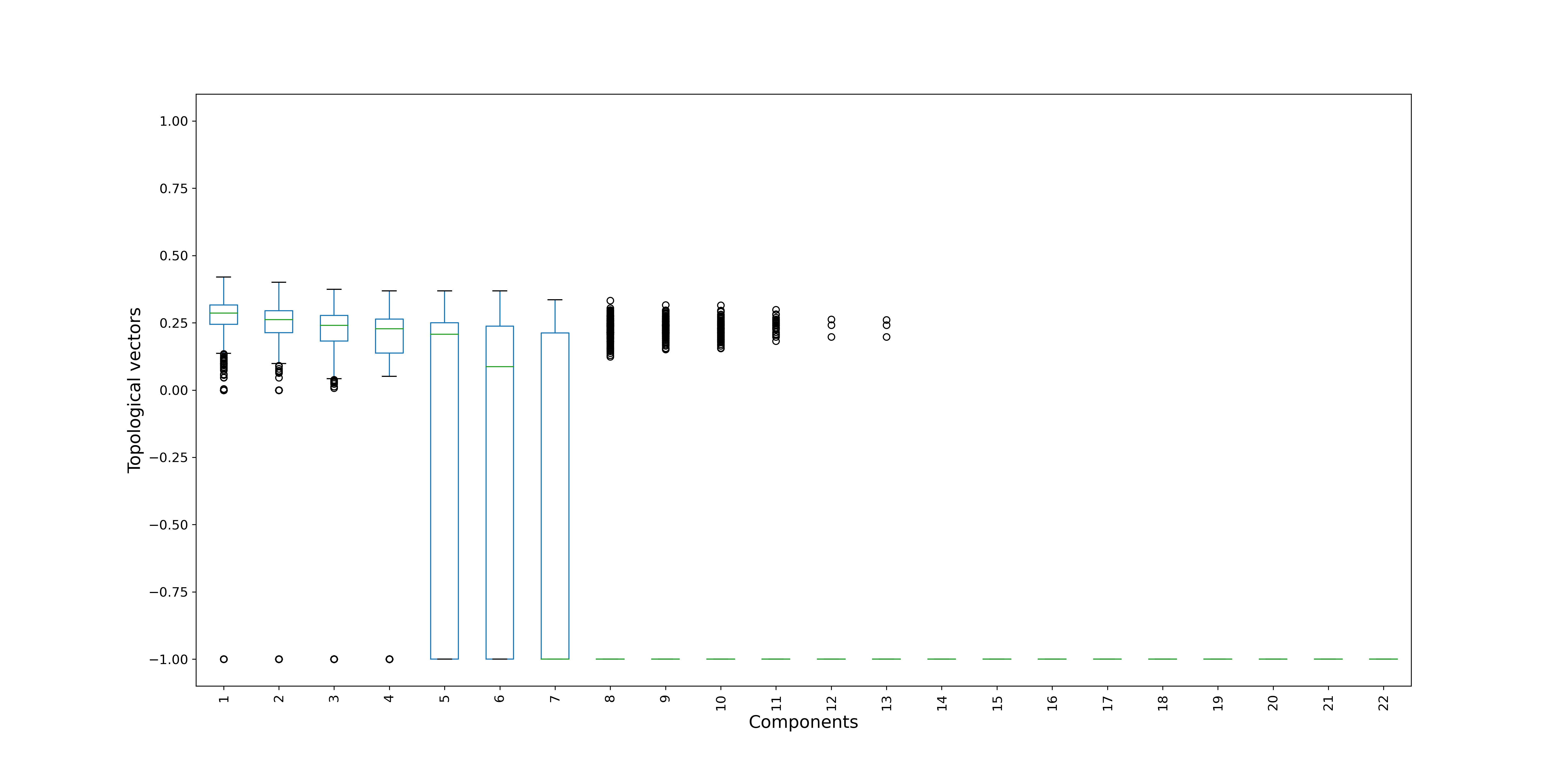}
	\caption{\label{c4_2}Boxplots of the topological vectors in homological dimension $H_{2}$ for data in $C_{4}$.}
\end{figure}

\begin{figure}[H]
	\centering
	\includegraphics[width=1\textwidth]{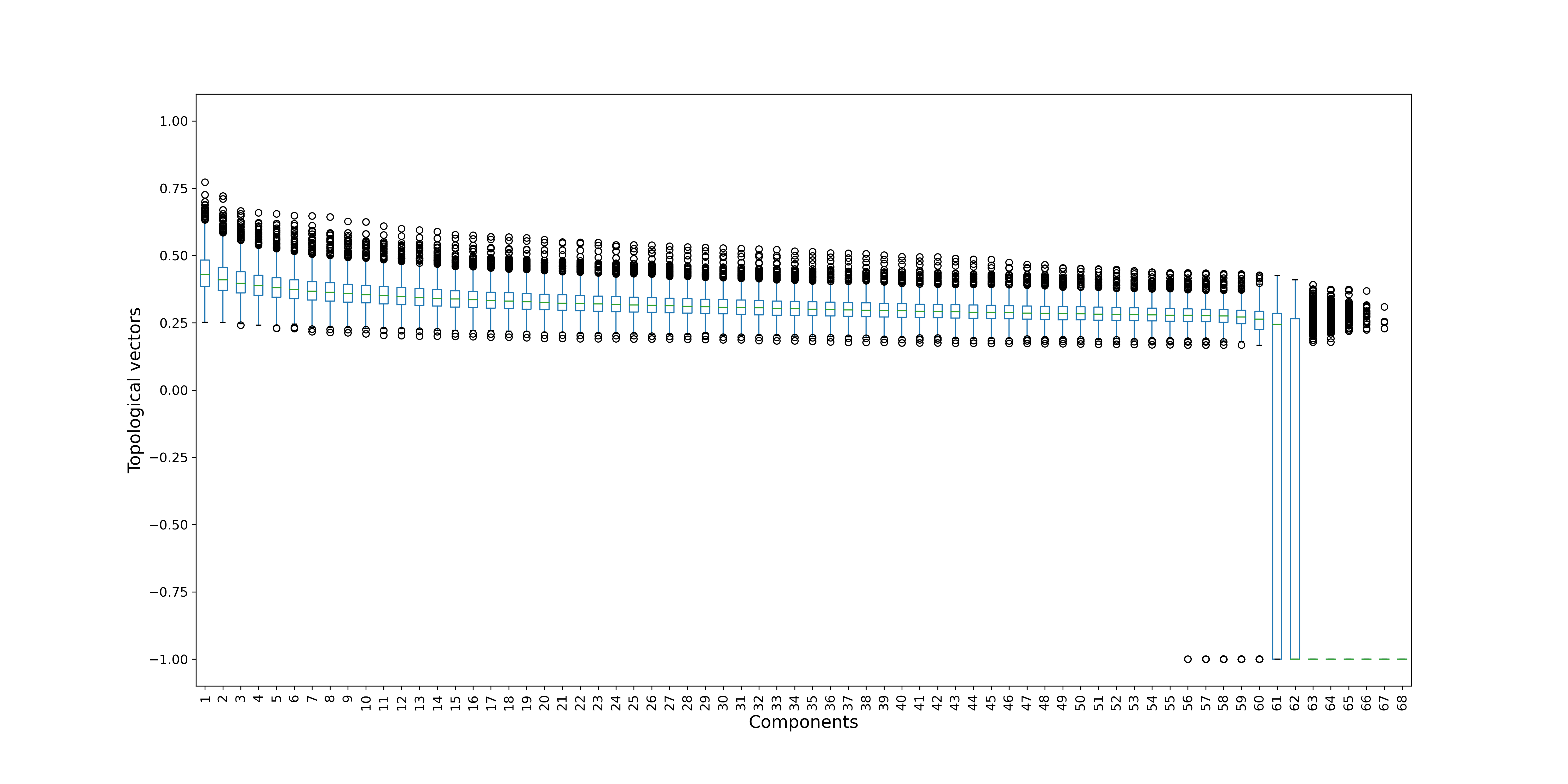}
	\caption{\label{c5_0}Boxplots of the topological vectors in homological dimension $H_{0}$ for data in $C_{5}$.}
\end{figure}

\begin{figure}[H]
	\centering
	\includegraphics[width=1\textwidth]{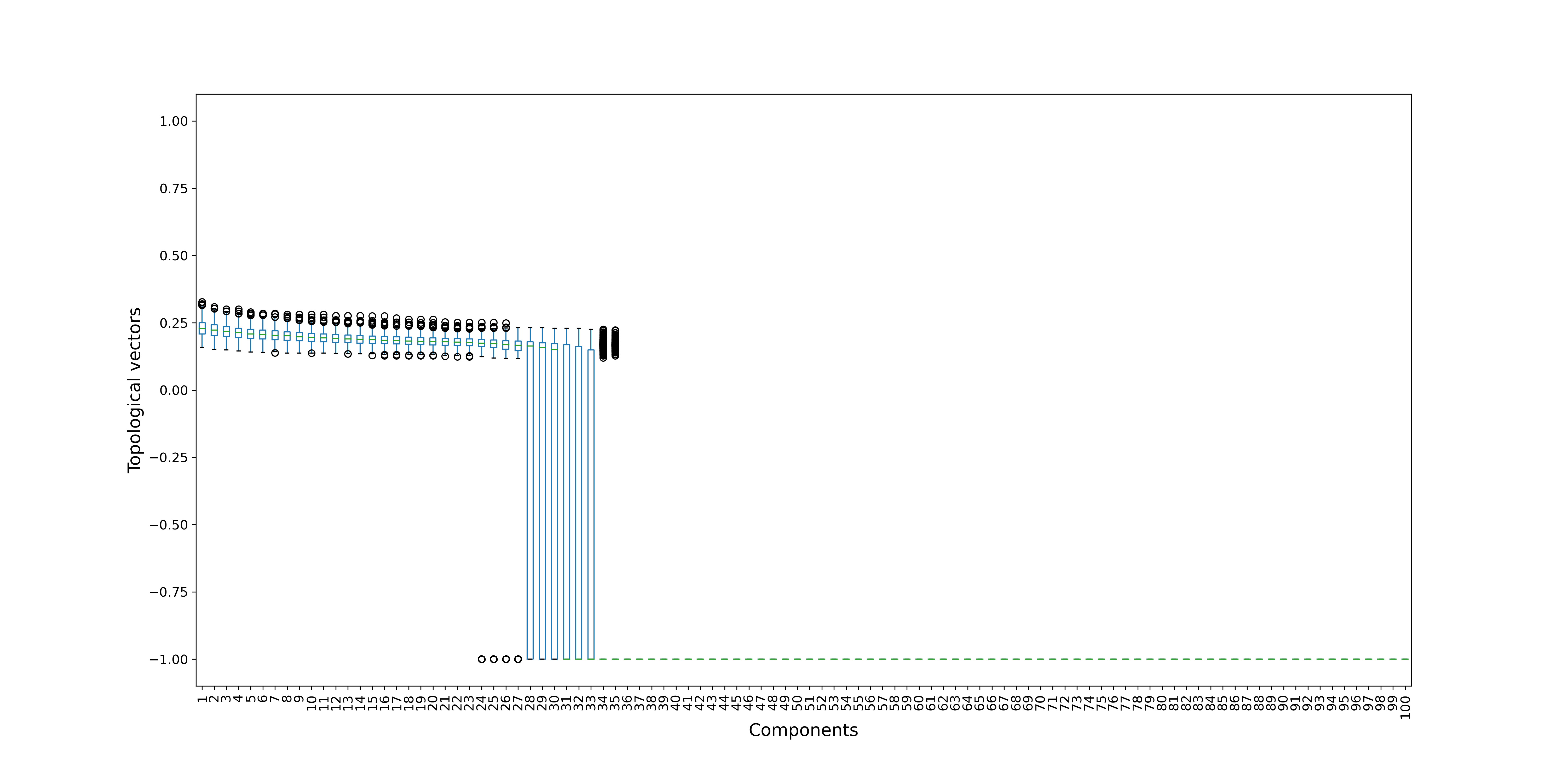}
	\caption{\label{c5_1}Boxplots of the topological vectors in homological dimension $H_{1}$ for data in $C_{5}$.}
\end{figure}

\begin{figure}[H]
	\centering
	\includegraphics[width=1\textwidth]{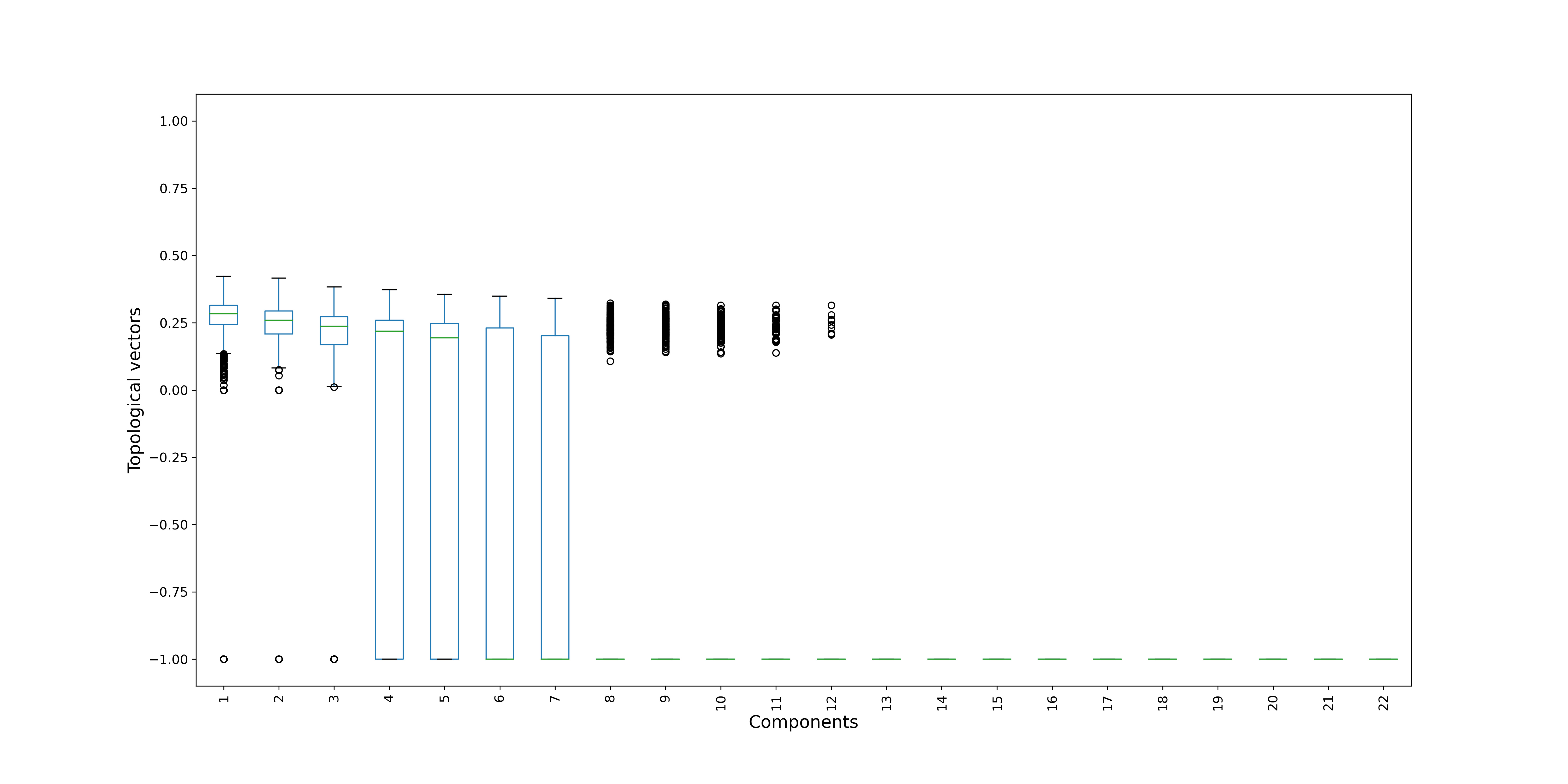}
	\caption{\label{c5_2}Boxplots of the topological vectors in homological dimension $H_{2}$ for data in $C_{5}$.}
\end{figure}

\begin{figure}[H]
	\centering
	\includegraphics[width=1\textwidth]{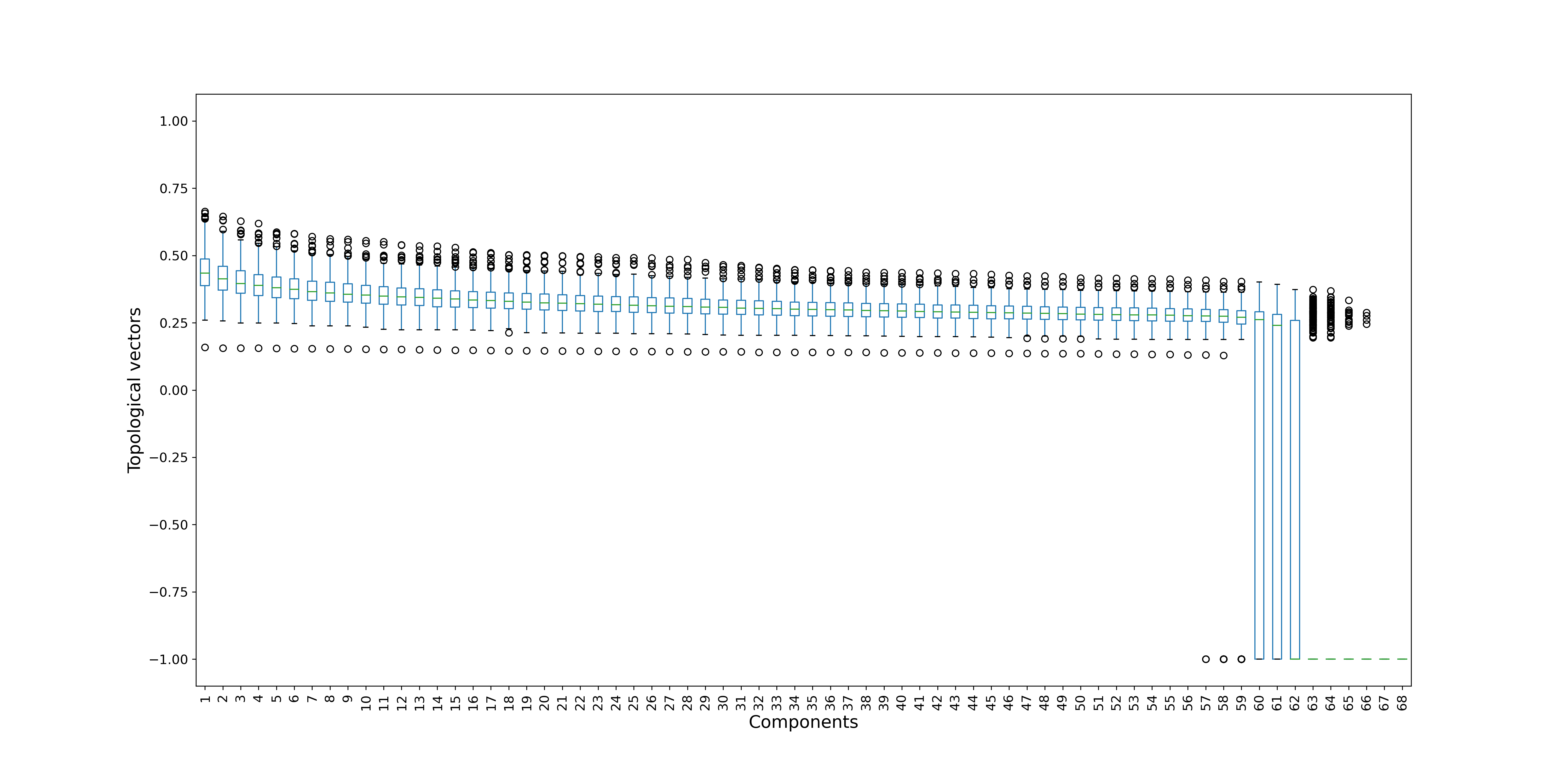}
	\caption{\label{c6_0}Boxplots of the topological vectors in homological dimension $H_{0}$ for data in $C_{6}$.}
\end{figure}

\begin{figure}[H]
	\centering
	\includegraphics[width=1\textwidth]{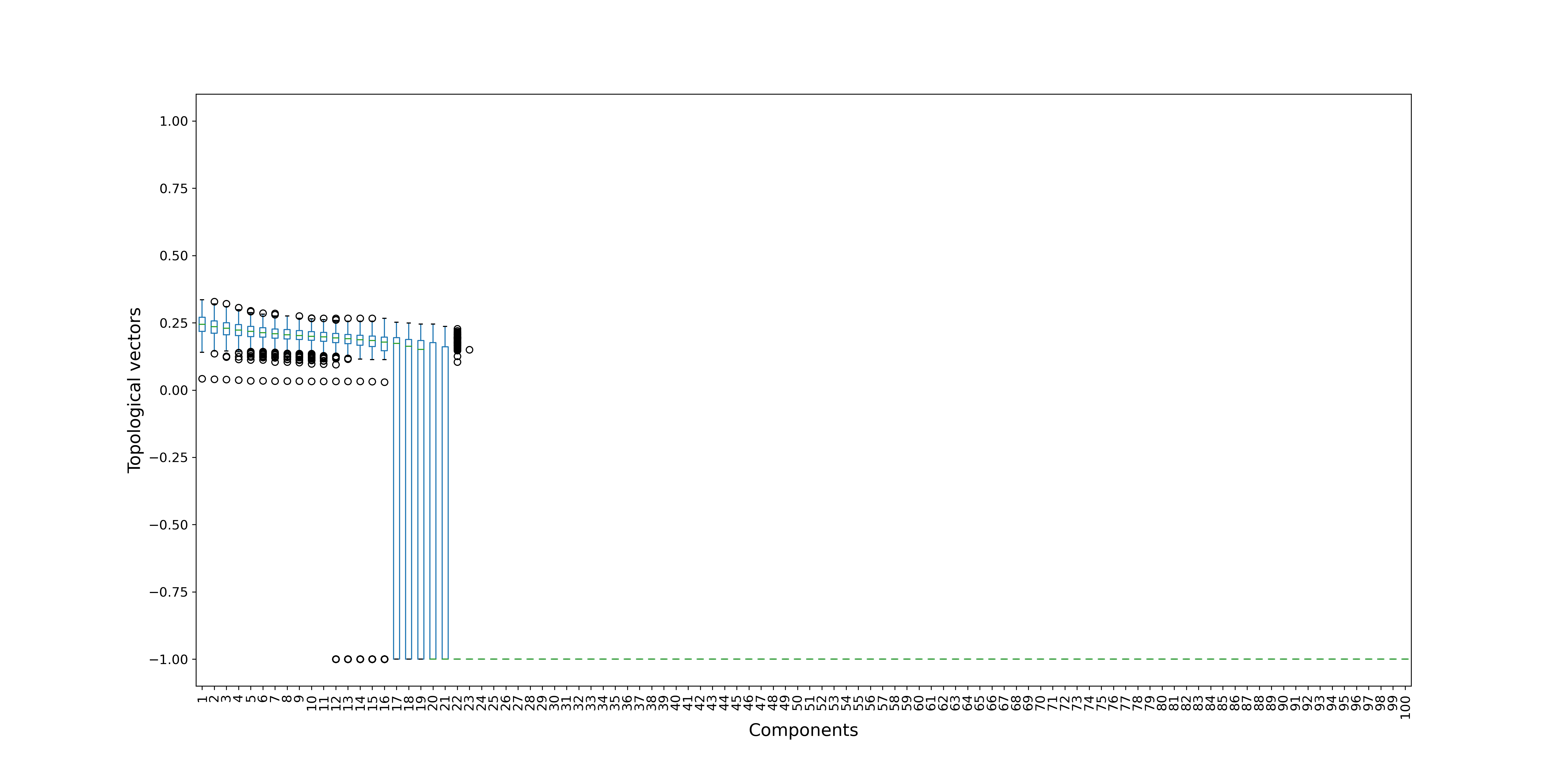}
	\caption{\label{c6_1}Boxplots of the topological vectors in homological dimension $H_{1}$ for data in $C_{6}$.}
\end{figure}

\begin{figure}[H]
	\centering
	\includegraphics[width=1\textwidth]{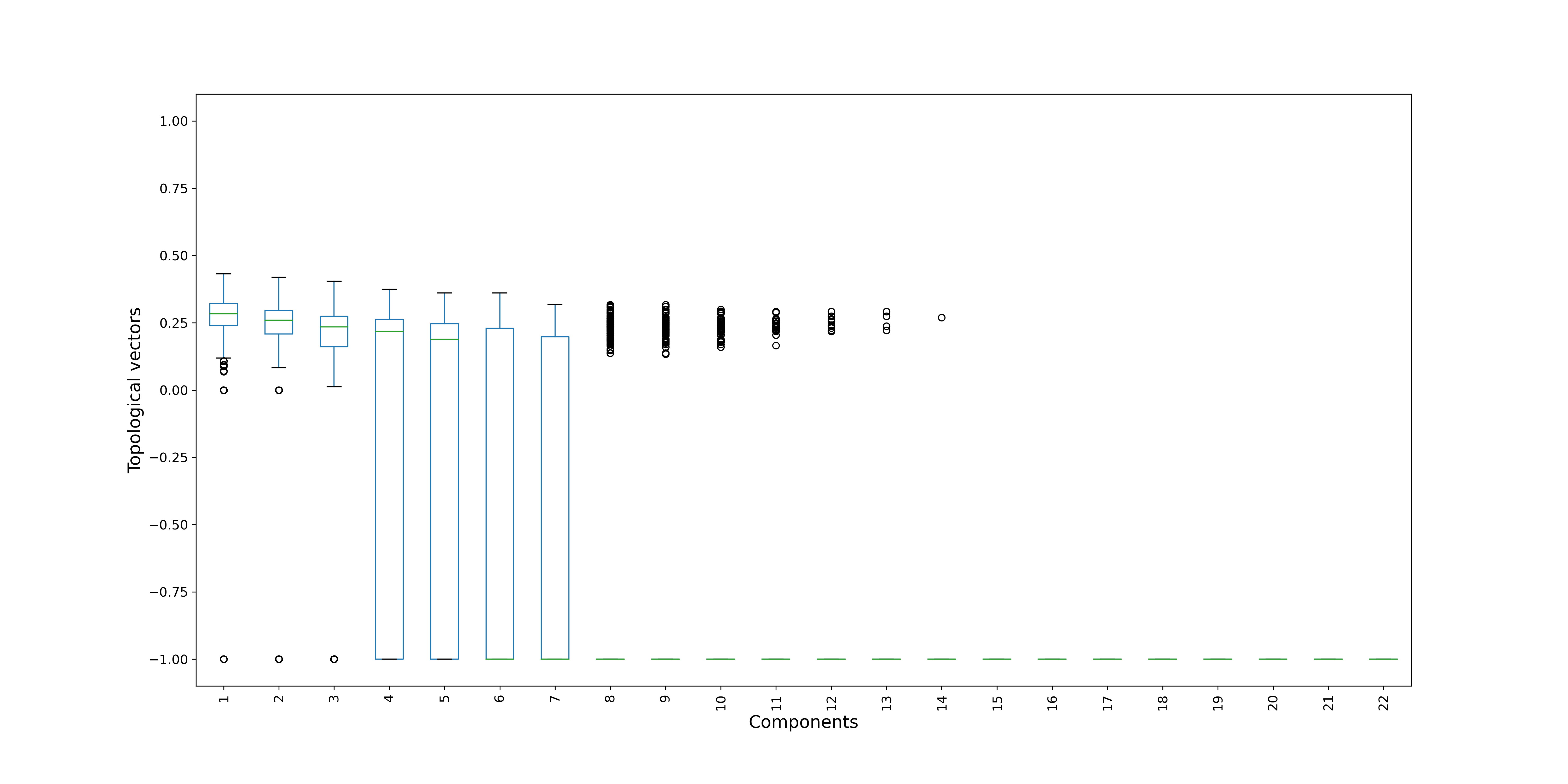}
	\caption{\label{c6_2}Boxplots of the topological vectors in homological dimension $H_{2}$ for data in $C_{6}$.}
\end{figure}

\begin{figure}[H]
	\centering
	\includegraphics[width=1\textwidth]{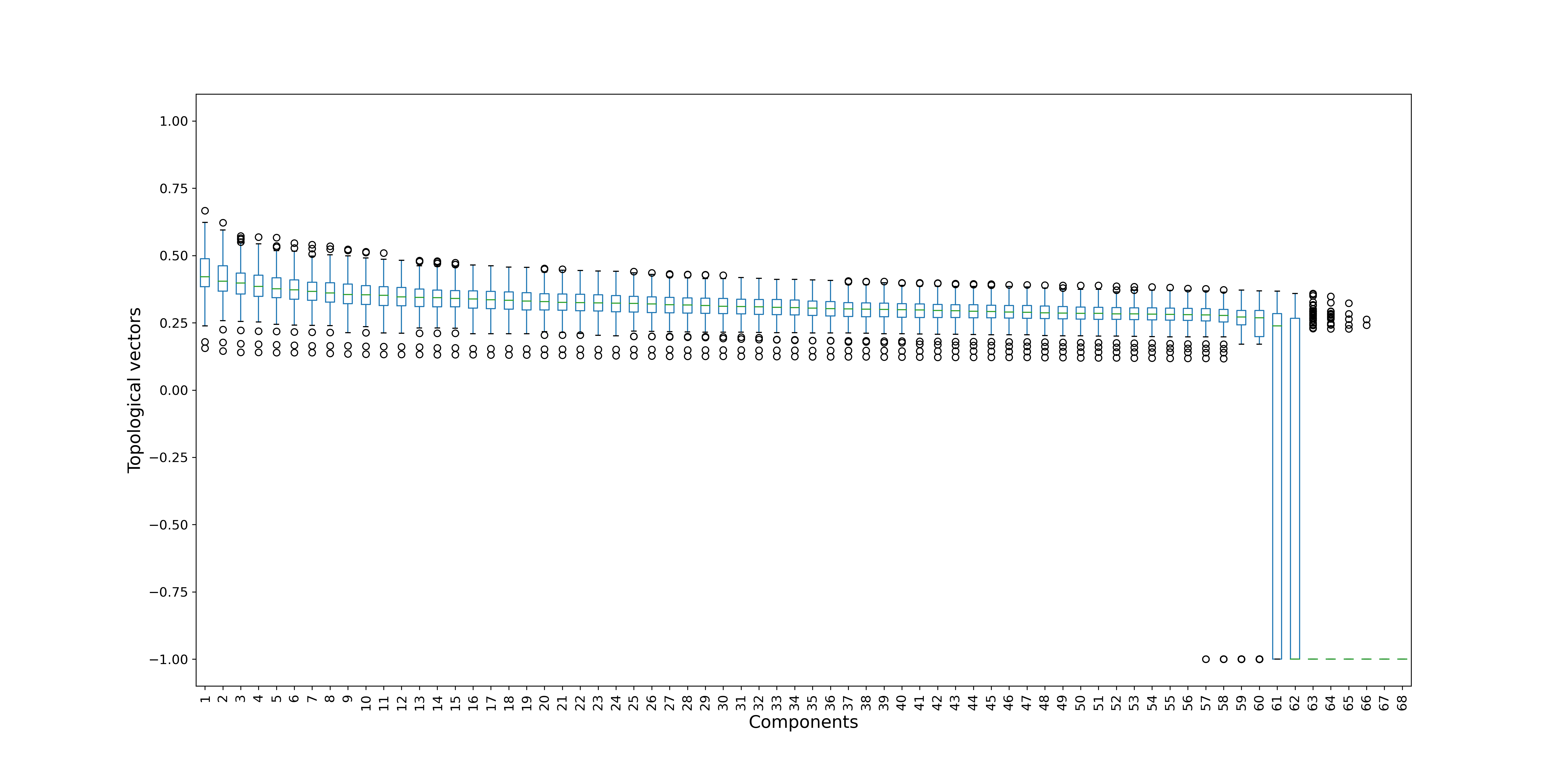}
	\caption{\label{c7_0}Boxplots of the topological vectors in homological dimension $H_{0}$ for data in $C_{7}$.}
\end{figure}

\begin{figure}[H]
	\centering
	\includegraphics[width=1\textwidth]{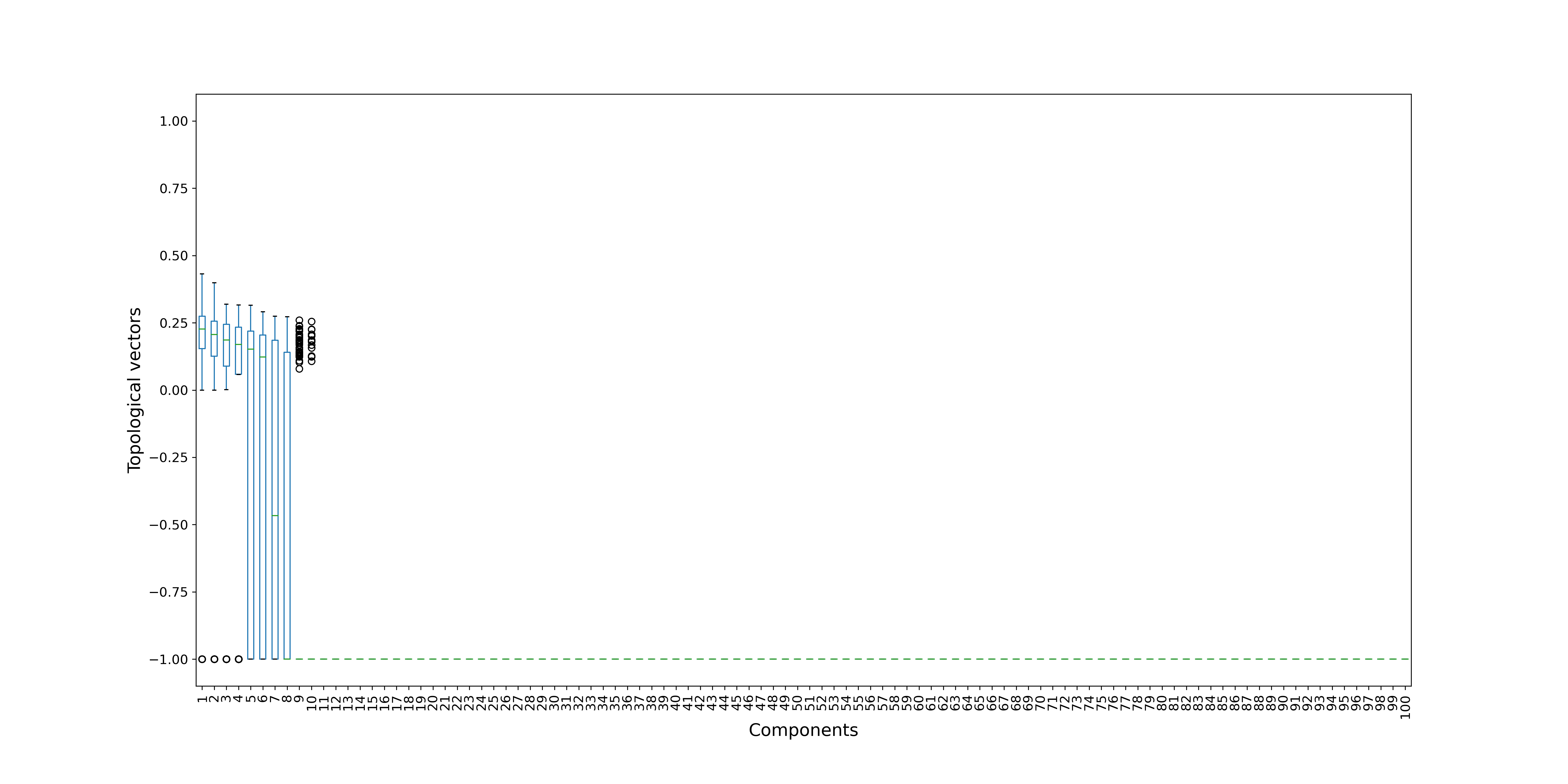}
	\caption{\label{c7_1}Boxplots of the topological vectors in homological dimension $H_{1}$ for data in $C_{7}$.}
\end{figure}

\begin{figure}[H]
	\centering
	\includegraphics[width=1\textwidth]{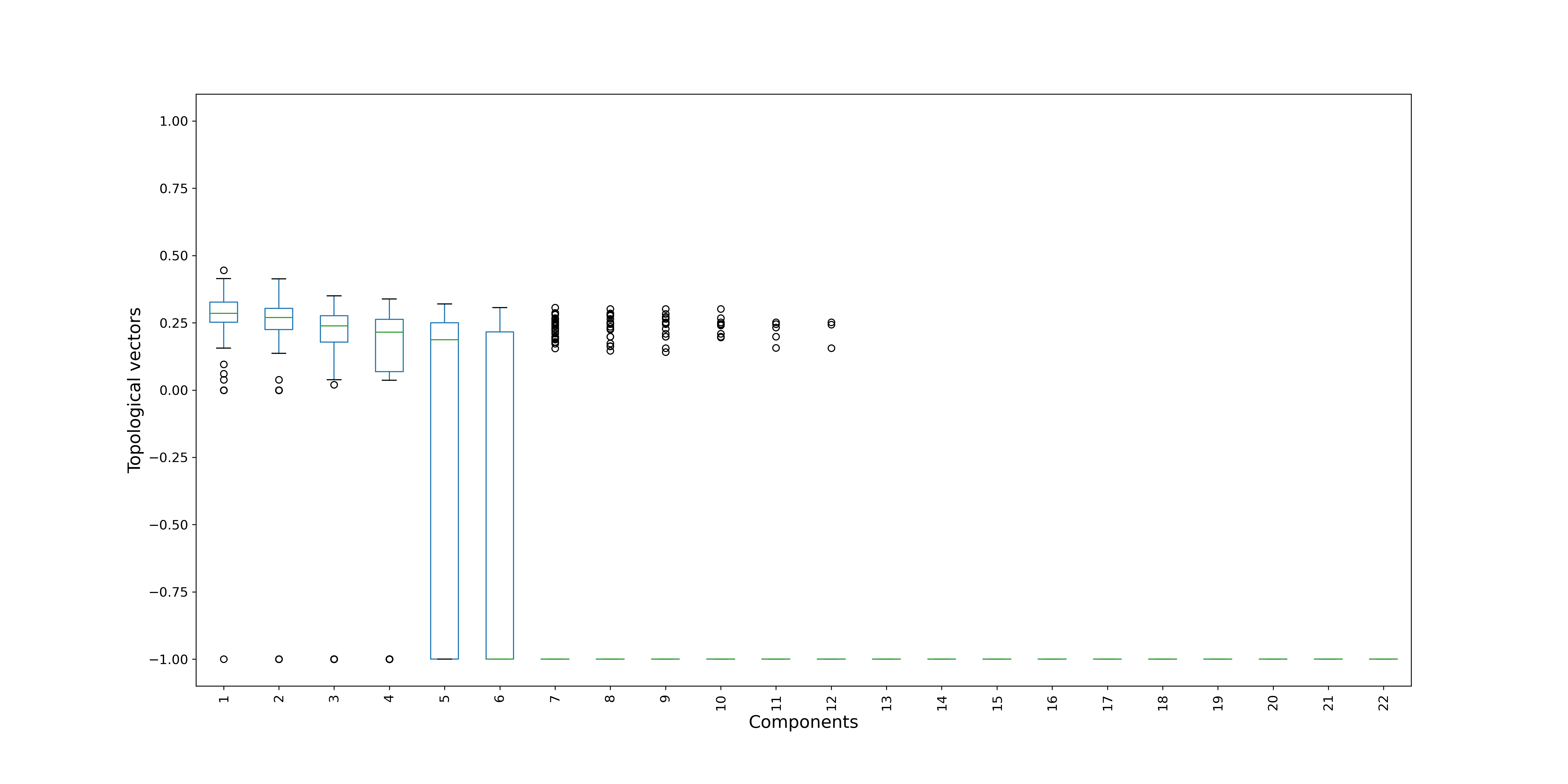}
	\caption{\label{c7_2}Boxplots of the topological vectors in homological dimension $H_{2}$ for data in $C_{7}$.}
\end{figure}

We show in the Figure~\ref{c1_0} to Figure~\ref{c7_2} below boxplots of the topological vectors in the homological dimensions $H_{0}$, $H_{1}$, and $H_{2}$ for each cluster $C_{i}$.

\subsection*{Cluster 1}

\subsection*{Cluster 2}

\subsection*{Cluster 3}

\subsection*{Cluster 4}

\subsection*{Cluster 5}

\subsection*{Cluster 6}

\subsection*{Cluster 7}

\section*{Embedded data representations}

\begin{figure}[!h]
	\centering
	\includegraphics[width=1\textwidth]{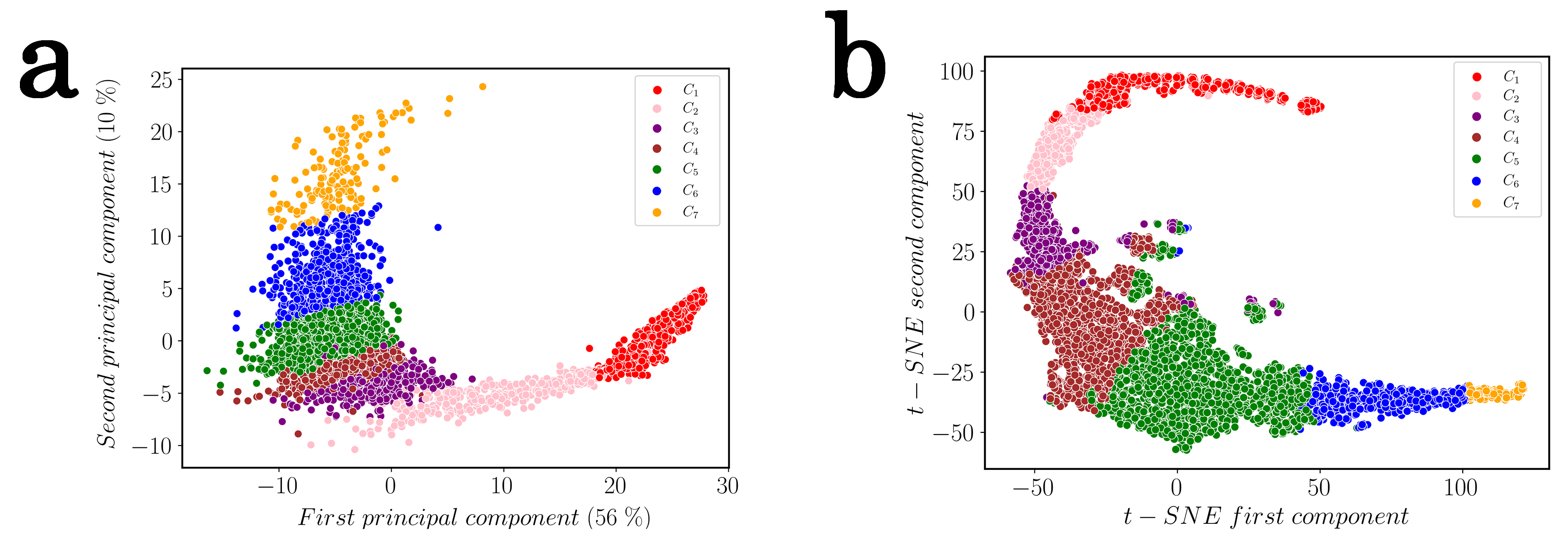}
	\caption{\label{fig:EmbeddedDataRepresentations}(a) PCA and (b) t-SNE first components of the trained model in the descriptor space.}
\end{figure}

The most interesting in-depth that we were able to get to interrogate visually the data was obtained by a classical Principal Component Analysis (PCA) \cite{Jolliffe2002} represented on Figure~\ref{fig:EmbeddedDataRepresentations}(a). The dissociation between solid and liquid-like structure is obvious on the first axis. However, there is no clear frontier between all the clusters due to the dimensionality reduction into such a plane (whereas our clusters are well separated, as given by the posterior probabilities). A similar graph is obtained on Figure~\ref{fig:EmbeddedDataRepresentations}(b) by a t-SNE algorithm \cite{Maaten2008} until convergence (around 20000 iterations) with a perplexity value of 75. The representation onto two axis is not perfect (66 \% of information is covered by PCA) but we can see that the width is different for each cluster ($C_1$ and $C_2$ being more diverse in the first axis) and also different volumes ($C_7$ seems to have a high variance for few points).

\section*{Bond-Orientational Order Analysis (BOOA)}

\begin{figure}[!h]
	\centering
	\includegraphics[width=1\textwidth]{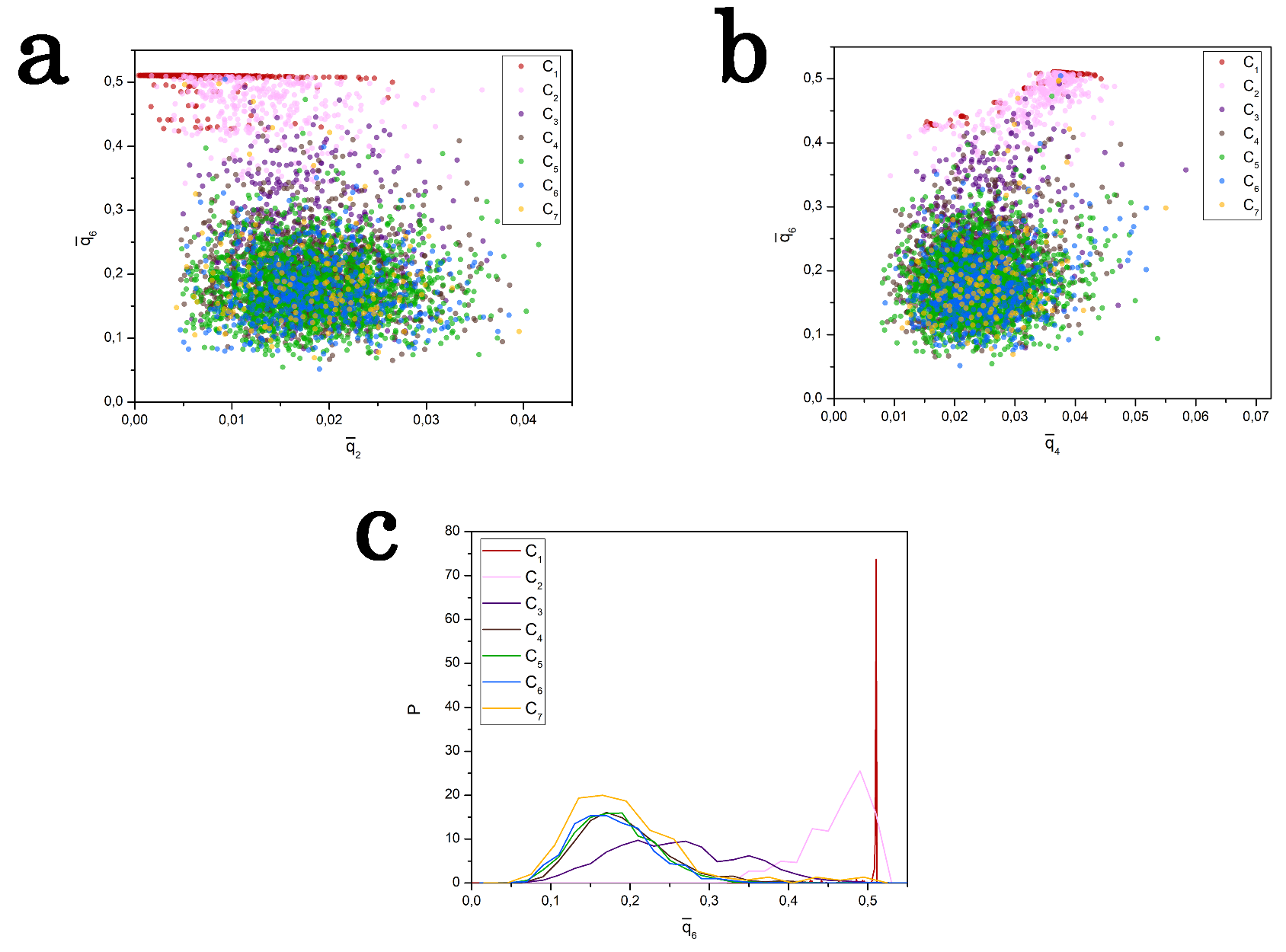}
	\caption{\label{fig:BOOA}(a) $\bar{q}_2$-$\bar{q}_6$-plane, (b) $\bar{q}_4$-$\bar{q}_6$-plane and (c) probability distributions of $\bar{q}_6$, for the structures in the trained model.}
\end{figure}

Among the set of parameters based on spherical harmonics which can be computed from the Bond-Orientational Order Analysis (BOOA), some of them are widely used in the literature such as the $\bar{q}_2$-$\bar{q}_6$-plane for identification of cristallinity, the $\bar{q}_4$-$\bar{q}_6$-plane for distinction of crystal structure or the $\bar{q}_6$ distribution for identification of solidity \cite{Dellago2008}. Figure~\ref{fig:BOOA} shows these representations on the trained model. The distinction between the structures associated with the liquid and solid clusters is highlighted. Figure~\ref{fig:BOOA}(c) is particularly informative on the fact that the structures in $C_1$ present a well-defined crystalline order, while the distribution in $C_2$ is more spread out. $C_3$ shows a boundary distribution between the clusters classified as liquid and the cluster $C_2$ mainly located at the border of the nuclei. These conclusions are in agreement and equivalent to our results obtained with the CNA. However, although the liquid clusters are identifiable with BOOA, it is not clear that the associated structures also possess a relative crystalline order. On the contrary, the CNA clearly reveals the existence of $[666]$ and $[444]$ bcc geometrical bonds in the structures of these liquid clusters. This feature of the CNA allowed us to investigate the heterogeneous scenario of nucleation as described in the manuscript.

\end{document}